\documentclass[showpacs,twocolumn,preprintnumbers,amsmath,amssymb,pra]{revtex4-2}
\usepackage{graphicx}
\usepackage{nicefrac}
\usepackage{color}
\usepackage[squaren]{SIunits}
\usepackage[normalem]{ulem}
\usepackage[usenames,dvipsnames]{xcolor}

\usepackage[pdftex,unicode,colorlinks,
]{hyperref}

\begin{document}
\title{Quantum spatial dynamics of high-gain parametric down-conversion accompanied by cascaded up-conversion}

\author{A.V. Rasputnyi}
\email{andrey@shg.ru}
\author{D.A. Kopylov}
\affiliation{Physics Department, Lomonosov Moscow State University, 119991 GSP-1, Leninskie Gory, Moscow, Russia}

 \begin{abstract}

Quantum cascaded up-conversion (CUpC) of parametric down conversion (PDC) in a finite nonlinear $\chi^{(2)}$-crystal is studied theoretically within parametric approximation.
The exact solution for creation and annihilation operators presented in the form of Bogoliubov transformation is valid for the high-gain regime and explicitly includes the non-zero wavevector-mismatch both for the PDC and CUpC.
With the use of characteristic equation parametric amplification and  oscillating regimes are analysed for degenerate, three- and four-mode cases.
We show that the parametric amplification exists under the fulfilment of the cascaded phase-matching conditions while both the PDC and CUpC processes are separately non-phase-matched.
The influence of CUpC on quadrature squeezing of degenerate PDC is estimated.

\end{abstract}

\maketitle

\section{Introduction}

Quantum light sources and frequency converters based on nonlinear optical effects are of special interest in quantum optics and technologies~\cite{klyshko1988book,Brecht_2015,Fabre_2020}.
Nowadays, the most widespread non-classical light sources are based on the second-order nonlinear effect of parametric down-conversion (PDC)~\cite{klyshko1988book}.
PDC is the parametric amplification of electromagnetic vacuum fluctuations that occurs in nonlinear crystals and leads to photon pairs creation in so-called signal and idler modes.
On the output of the crystal the squeezed vacuum state~\cite{Walls_1983}(or the bright squeezed vacuum state in the high-gain regime~\cite{Chekhova_2015}) is realized.
Signal and idler photons reveal quantum correlations that allows one e.g. to prepare entangled states~\cite{Shih_2003,Fabre_2020} and overcome shot-noise limit~\cite{LIGO_2013,Vahlbruch_2016}.

Simultaneously with PDC, signal and idler photons can be involved in the same nonlinear crystal into additional nonlinear processes, called `cascaded' or `multistep' processes~\cite{Saltiel_2005}.
One of such processes is the cascaded up-conversion (CUpC) of signal (or idler) PDC photons, which leads to coupling of four modes: signal, idler and their up-converted modes.
The CUpC of PDC is known also as the `cascaded hyperparametric scattering'~\cite{klyshko1988book} or as the `the parametric amplification at low-frequency pump'~\cite{Akhmanov_1972_book}, when the seed in signal (or idler) wave is present.

The PDC with CUpC is of interest as nonclassical light source with unique properties. 
The first observations of CUpC from PDC were obtained in the 1970-th for the three-mode interaction, when CUpC process took place for the signal (or idler) PDC mode~\cite{Andrews_1970,Chirkin_1974}, and the nonclassical statistical properties of the generated light were studied theoretically~\cite{Tang_1969,Mishkin_1969,Klyshko_1970,Smithers_1974,Il_inskii_1975}.
Later the tri-partite entangled states based on the CUpC were implemented~\cite{Ferraro_2004,Puddu_2004,Allevi_2006,Allevi_2008} and the spectral properties of the broadband CUpC were experimentally studied~\cite{Sun_2009}. 
In addition to the three-mode case, the degenerate regime (signal and idler modes are not distinguishable)~\cite{Perina_1995_2,Chirkin_2002}, and the four-mode generation (both the signal and idler waves are up-converted)~\cite{Tlyachev_2014,Arkhipov_2016} were considered.

The effect of PDC with CUpC arises also in the quantum frequency converters (QFC) that are used for the detection of IR-radiation at the single photon level~\cite{Mancinelli_2017,Barh_2019} or for the entanglement support between remoted ions~\cite{Krutyanskiy_2019}. 
The PDC is present in QFC as a fundamental noise that was demonstrated e.g. in~\cite{Pelc_2010,R_tz_2017,Maring_2018,Strassmann_2019}.

The common quantum description of PDC with CUpC is realized in terms of the temporal evolution~\cite{Smithers_1974,Il_inskii_1975,Allevi_2006,Chirkin_2002,Tlyachev_2013,Tlyachev_2014,Arkhipov_2016}, while the non-zero wavevector-mismatch for considered nonlinear processes are omitted or involved effectively into the coupling constants with the use of short-length crystal approximation. 
However, the spatial dynamics and correct accounting of non-zero wavevector-mismatches can be critical for the generation of quantum light via nonlinear optical effects e.g. for the broadband multimode high-gain PDC~\cite{Christ_2013,Sharapova_2020}.

In contrast to the temporal evolution, the dynamics of the quantized electromagnetic field inside nonlinear crystals can be described in terms of the spatial evolution~\cite{Shen_1967,Huttner_1990}. 
This approach was successfully applied to the PDC generation (e.g. Refs.~\cite{Huttner_1990,Pe_ina_2015,Lipfert_2018,Horoshko_2019}), analysis of optical harmonics generation from multimode broadband PDC~\cite{Dayan_2007_theory,Kopylov_2020} and the investigation of the properties of the quantum nonlinear couplers~\cite{Perina_1995_2,Thapliyal_2014}.
To the best of our knowledge the spatial dynamics along the nonlinear crystal of the PDC with CUpC has not been previously considered.

In this paper we apply the formalism of the spatial evolution of the quantized light to the PDC accompanied by CUpC.
The paper is organized as follows: in Section~\ref{sec_approach} we consider the main aspects of the studied nonlinear optical processes and reduce the initial Heisenberg equations for annihilation operators to the ordinary differential system for the Bogoliubov functions that is solved analytically.
In Section~\ref{sec_deg_case} the degenerate PDC with CUpC is studied and the oscillating regimes and parametric amplification are analysed.
In addition the influence of CUpC on PDC squeezing properties is considered.
In Section~\ref{sec_nondeg_case} our approach is applied to the three- and four-mode CUpC of PDC.
The parametric amplification for cascaded phase-matching conditions is demonstrated.

\section{Theoretical approach}
\label{sec_approach}

For the quantum description of coupled PDC with CUpC in the transparent one dimensional finite nonlinear crystal we use the formalism based on momentum operator of the electromagnetic field~\cite{Shen_1967,Huttner_1990}.
In Heisenberg representation the light that propagates along the $z$-axis inside the dispersive medium is presented in terms of discrete temporal modes, with the frequencies $\omega_m=2\pi m /T$, where $m = 0, 1, 2\dots$ and $T$ -- quantization time. 
The electric field operator (the polarization indexes are omitted) has the form
\begin{equation}
	\hat{E}(z, t) = \sum_{\omega} \sqrt{\dfrac{\hbar \omega}{2\epsilon_0 c T n(\omega)} } \bigg( \hat{f}_{\omega}(z) e^{-i\omega t} + \hat{f}_{\omega}^\dagger(z) e^{i\omega t} \bigg),
\end{equation}
where $\hat{f}_{\omega}(z)$ and $\hat{f}_{\omega}^\dagger(z)$ are annihilation and creation operators with the bosonic commutation relations
\begin{align}\begin{split}
 [\hat{f}_{\omega}(z), \hat{f}^\dagger_{\omega^\prime}(z) ] &= \delta_{\omega \omega^\prime}, \\
[\hat{f}_{\omega}(z), \hat{f}_{\omega^\prime}(z) ] &= [\hat{f}^\dagger_{\omega}(z), \hat{f}^\dagger_{\omega^\prime}(z) ] = 0.
 \label{eq_commutation}
\end{split}\end{align}

The annihilation operators satisfy the Heisenberg equation~\cite{Huttner_1990}
\begin{equation}
    \dfrac{d \hat{f}(z)}{dz} =  \dfrac{i}{\hbar} [\hat{f}(z), \hat{G}(z)],  
    \label{eq_heisenberg_eq}
\end{equation}
where the momentum operator $\hat{G}(z)$ is the generator for the spatial evolution.

In this paper we assume that the state on the input of the nonlinear crystal is vacuum for all modes, thus the quantum-mechanical averaging $\langle ... \rangle$ is obtained over the vacuum state $|0\rangle$. 
So the mean number of photons $\mathcal{N}_{f\omega}(z) $ in the mode $\hat{f}_{\omega}(z)$ has the form
\begin{equation}
    \mathcal{N}_{f\omega}(z) \equiv \langle 0|\hat{f}_{\omega}^{\dagger}(z)\hat{f}_{\omega}(z)|0\rangle.
    \label{eq_spectra}
\end{equation}

In addition to the mean number of photons the squeezing properties of interacting modes are studied. 
For the arbitrary collective mode
\begin{equation}
  \hat{F}(\delta, z) = \dfrac{\hat{f}_{\omega_1}(z) + e^{i\delta}\hat{f}_{\omega_2}(z)}{\sqrt{2}}, 
  \label{eq_collective_mode}
\end{equation}
the quadrature operator
\begin{equation}
    \hat{X}_{F}(\theta,\delta, z) = \hat{F}(\delta, z) e^{i\theta} + \hat{F}^{\dagger}(\delta, z) e^{-i\theta}.
    \label{eq_quadrature}
\end{equation}
can be introduced.
For $\theta=0$ and $\theta=\pi/2$ this operator corresponds to the `position' $\hat{Q}_{F}$ and `momentum' $\hat{P}_{F}$ quadratures, respectively.
The variance of the quadrature~\eqref{eq_quadrature} has the form
\begin{equation}
 (\Delta X_{F}(\theta,\delta, z))^2  = \langle \hat{X}^2_{F}(\theta,\delta, z) \rangle    - \langle \hat{X}_{F}(\theta,\delta, z)^2 \rangle.
 \label{eq_variance}
\end{equation}

\begin{figure}[t]
\begin{center}
\includegraphics[width=0.8\linewidth]{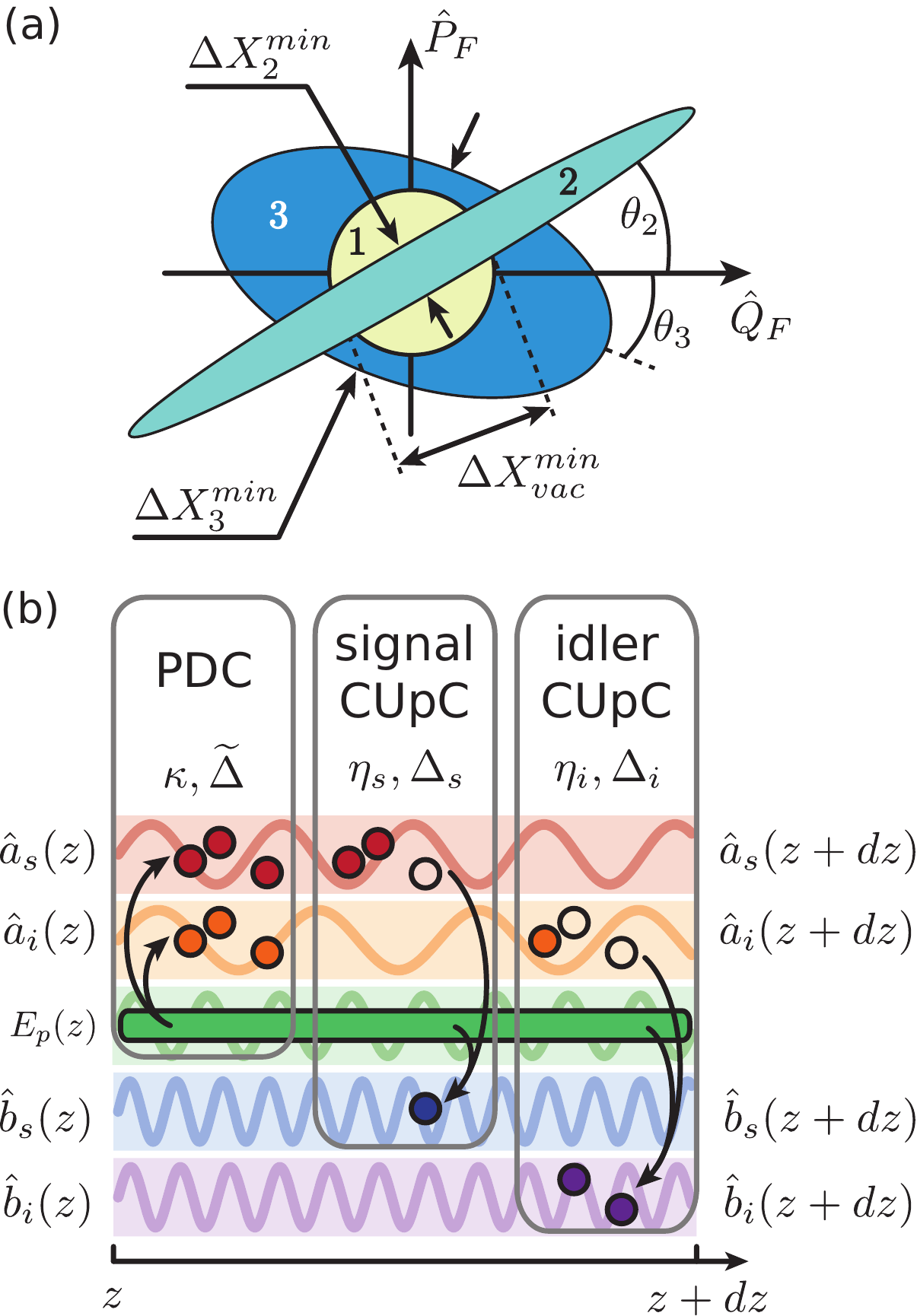}
\caption{
(a) Phase space plot, showing the uncertainty for 3 different states: 1 -- vacuum state; 2 -- squeezed state with $(\Delta X_2^{min})^2<(\Delta X_{vac}^{min})^2$; 3 -- state with $(\Delta X_3^{min})^2>(\Delta X_{vac}^{min})^2$. The minimal quadrature dispersion for state 2 and 3 is achieved at angles $\theta_2$ and $\theta_3$, respectively.
(b) Schematic representation of PDC with CUpC. At each point of the nonlinear crystal three simultaneous processes occur: PDC (photons are created in signal and idler modes), cascaded up-conversion of signal photons and idler photons. 
}
\label{fig_1_scheme}
\end{center}
\end{figure}

The quadrature variance~\eqref{eq_variance} depends on two parameters~($\theta,\delta$) and its minimal value $(\Delta {X}_{F}^{min})^2$ can be used as the characteristic of the squeezing properties.
For the vacuum state the minimal variance of quadrature~$(\Delta {X}_{vac}^{min})^2=1$ and does not depend on the angles~($\theta,\delta$).
For the arbitrary quantum state the minimal variance~$(\Delta {X}^{min})^2$ can be either larger (anti-squeezed state) or lower (squeezed state) than the vacuum one (Fig.~\ref{fig_1_scheme}(a)).

In the case of quadrature squeezing of single mode~$\hat{f}_{\omega}$ it is sufficient to consider the quadrature in the form $\hat{X}_{f}(\theta, z) = \hat{f}_{\omega} e^{i\theta} +\hat{f}_{\omega}^{\dagger}(z) e^{-i\theta}$.

\subsection{Spatial dynamics of PDC with CUpC}

Our description of PDC with CUpC is based on several assumptions.
Firstly, the parametric approximation is applied, therefore a classical monichromatic wave in the form $E_{p} = \frac{1}{2} \mathcal{E} e^{-i(\omega_p t - k_p z)} + \text{c.c.}$ is used as a pump wave, where $\mathcal{E}$ is the complex amplitude of the field, $\omega_p$ -- pump frequency, $k_p = k(\omega_p) $ -- pump wavevector in the nonlinear crystal. 
Secondly, we omit all the effects caused by the polarization of the light: the coupling of the interacting modes is described by effective susceptibilities.

Three simultaneous second order processes are considered in the nonlinear crystal: PDC ($\omega_p \rightarrow \omega_{as}+\omega_{ai}$), signal CUpC $\omega_p+\omega_{as} \rightarrow \omega_{bs}$, idler CUpC ($\omega_p+\omega_{ai} \rightarrow \omega_{bi}$).
Here and further in the text the indexes $a$ and $b$ correspond to the PDC and CUpC modes, respectively, and indexes $s$ and $i$ --- signal and idler waves.

Thereby four quantized modes are involved into the nonlinear interaction (Fig.~\ref{fig_1_scheme}(b)): 
\begin{align} 
 \hat{a}_s(z) &\equiv \hat{f}_{\omega_{as}}(z)      = \hat{\alpha}_s(z)e^{ik_{as} z}    & \text{(PDC signal)},\\
 \hat{a}_i(z) &\equiv \hat{f}_{\omega_{ai}}(z)      = \hat{\alpha}_i(z)e^{ik_{ai} z}    & \text{(PDC idler)}, \\
 \hat{b}_s(z) &\equiv \hat{f}_{\omega_{bs}}(z) = \hat{\beta}_s(z)e^{ik_{bs} z}    & \text{(CUpC signal)}, \\
 \hat{b}_i(z) &\equiv \hat{f}_{\omega_{bi}}(z) = \hat{\beta}_i(z)e^{ik_{bi} z}    & \text{(CUpC idler)}.
 \label{eq_sig_idl_operators}
 \end{align}
where the slowly-varying annihilation operators $\hat{\alpha}(z)$ and $\hat{\beta}(z)$ are introduced, and $k_{n}=k(\omega_n)$ are the wavevectors for each mode in nonlinear medium.

The momentum operator for studied system in Heisenberg representation consists of two terms $\hat{G}(z) = \hat{G}_l(z) + \hat{G}_{nl}(z)$.
The linear part $\hat{G}_l(z)$ describes the propagation for each mode in the linear dielectric medium
\begin{equation}
\hat{G}_l(z) =  \dfrac{1}{2} \sum_{{f}= {a},{b}} \sum_{n=s,i} \hbar k_{f n} \hat{f}_n^\dagger(z)\hat{f}_n(z) + h.c. 
\label{eq_momentum_linear}
\end{equation}
The nonlinear term $\hat{G}_{nl}$ characterizes the nonlinear interaction that is present as a sum of two contributions $\hat{G}_{nl}(z) = \hat{G}_{pdc}(z) + \hat{G}_{upc}(z)$, where 
\begin{equation}
  \hat{G}_{pdc}(z) = \hbar \kappa \ \hat{a}_s^\dagger(z) \hat{a}_i^\dagger(z)e^{i k_p z} + h.c.,
  \label{eq_momentum_pdc}
\end{equation}
corresponds to PDC (photons creation in signal and idler modes).
The momentum operator for signal ($s$) and idler ($i$) CUpC is following
\begin{multline}
  \hat{G}_{upc}(z) = \hbar \eta_s \ \hat{a}_s (z) \hat{b}_s^\dagger(z)e^{i k_p z} \\ 
  + \hbar \eta_i \ \hat{a}_i (z) \hat{b}_i^\dagger(z)e^{i k_p z} + h.c.
  \label{eq_momentum_upc}
\end{multline}
Here $\kappa \propto \bar\chi^{(2)}_{pdc} \ \mathcal{E} $, $\eta_{s,i} \propto \bar\chi^{(2)}_{s,i} \ \mathcal{E} $ are the complex coupling constants for PDC and for signal and idler CUpC, respectively.
The $\bar\chi^{(2)}_n$ are the effective nonlinear susceptibilities for each process.

By substitution~\eqref{eq_momentum_linear},~\eqref{eq_momentum_pdc},~\eqref{eq_momentum_upc} into~\eqref{eq_heisenberg_eq} and applying the expression $ d\hat{a}(z)/dz  = ik_{a} \hat{a}(z) +  e^{ik_{a} z} \ d\hat{\alpha}(z)/dz $ , the two independent systems of Heisenberg equations for slowly-varying operators are obtained.  
The first one is
\begin{equation}
    \begin{cases}
   \dfrac{d\hat{\alpha}_s(z)}{dz}  = i\kappa e^{i \widetilde\Delta z} \hat{\alpha}_i^{\dagger}(z)+ i\eta_s^* e^{ i \Delta_s z}\hat{\beta}_s(z),
   \\
   \dfrac{d\hat{\beta}_s(z)}{dz}  =  i\eta_s e^{-i \Delta_s z}\hat{\alpha}_s(z),
   \\
   \dfrac{d\hat{\alpha}^\dagger_i(z)}{dz}  = - i\kappa^* e^{-i \widetilde\Delta z} \hat{\alpha}_s (z) - i \eta_i  e^{-i\Delta_i z}\hat{\beta}_i^\dagger(z),
   \\
   \dfrac{d\hat{\beta}^\dagger_i(z)}{dz}  = - i\eta_i^*  e^{ i\Delta_i z}\hat{\alpha}^\dagger_i(z),
   \end{cases}
   \label{eq_svo_system}
\end{equation}
while the second one has the same form as~\eqref{eq_svo_system} but with the replaced indexes $i\leftrightarrow s$.
Here the $\widetilde\Delta = k_p - k_{as} - k_{ai}$ is the wavevector-mismatch of PDC and $\Delta_{s,i}=k_{bs,bi} - k_{as,ai} - k_p$ are the wavevector-mismatches of CUpC for signal ($s$) and idler ($i$) modes.
 
\paragraph{Exact solution}
\label{par_exact_solution}

In this paper the exact solution of the equations~\eqref{eq_svo_system} is found by using the approach presented in Ref.~\cite{Christ_2013}.
So far as the momentum operator has the bilinear form and the equations~\eqref{eq_svo_system} are linear on creation and annihilation operators, the solution can be presented in the form of Bogoliubov transformation (see Appendix~\ref{appendix_bogoliubov})

\begin{equation}
\begin{pmatrix} 
  \hat{\alpha}_{s}(z) \\
  \hat{\beta}_{s}(z) 
\end{pmatrix}
 = 
\begin{pmatrix} 
    U_{s}(z) & V_{s}(z) & W_{s}(z) & Q_{s}(z) \\
    K_{s}(z) & L_{s}(z) & M_{s}(z) & N_{s}(z) \\
\end{pmatrix}
\begin{pmatrix} 
  \hat{\alpha}_{s}(0) \\
  \hat{\alpha}_{i}^{\dagger}(0) \\
  \hat{\beta}_{s}(0) \\
  \hat{\beta}_{i}^{\dagger}(0)
\end{pmatrix}.
\label{eq_bogtransform_1}
\end{equation}
The similar transformation for the operators $\alpha_i(z)$ and $\beta_i(z)$ have the form~\eqref{eq_bogtransform_1} with replaced indexes $s \leftrightarrow i$.
 
Substituting~\eqref{eq_bogtransform_1} into the system \eqref{eq_svo_system} and combining the coefficients before the operators one can get two differential systems for the introduced Bogoliubov's functions
\begin{equation}
  \begin{cases}
 \dfrac{d U_s(z)}{dz}  =i \kappa \ e^{i\widetilde\Delta z} \ V_i^*(z)+ i \eta_s^* \ e^{i\Delta_s z} \ K_s(z) ,  
 \\
 \dfrac{d V_i^*(z)}{dz}  =-i \kappa^* e^{-i\widetilde\Delta z} U_s(z) - i \eta_i e^{-i\Delta_i z} L_i^*(z),
 \\
 \dfrac{d K_s(z)}{dz}  = i \eta_s \ e^{-i\Delta_s z} \ U_s(z),
 \\
 \dfrac{d L_i^*(z)}{dz}  = -i \eta_i^* \ e^{i \Delta_i z} \ V_i^*(z),
 \\
 U_s(0) = 1,  V_i^*(0)=0,  K_s(0)=0,  L_i^*(0) = 0,
 \end{cases}
 \label{eq_func_system_1}
 \end{equation}
\begin{equation}
  \begin{cases}
 \dfrac{d  W_s(z)}{dz} =i \kappa e^{i\widetilde\Delta z} \ Q_i^*(z)+ i \eta_s^* \ e^{i\Delta_s z} \ M_s(z),
 \\
 \dfrac{d Q_i^*(z)}{dz} =-i \kappa^* e^{-i\widetilde\Delta z} W_s(z) - i \eta_i e^{-i\Delta_i z} N_i^*(z),
 \\
 \dfrac{d  M_s(z)}{dz} = i \eta_s \ e^{-i\Delta_s z} \ W_s(z),
 \\
 \dfrac{d  N_i^*(z)}{dz} = -i \eta_i^* \ e^{i \Delta_i z} \ Q_i^*(z),
 \\
 W_s(0) = 0, Q_i^*(0)=0, M_s(0)=1, N_i^*(0) = 0.
\   \end{cases}
 \label{eq_func_system_2}
\end{equation}

The equations for the functions $U_i(z)$, $V_s^*(z)$, $K_i(z)$, $L_s^*(z)$ and $W_s(z)$, $Q_i^*(z)$, $M_s(z)$, $N_i^*(z)$ are similar to the equations~\eqref{eq_func_system_1} and \eqref{eq_func_system_2}, respectively, but with replaced indexes $i \leftrightarrow s$. 

Thus, the initial systems of differential equations for the operators~\eqref{eq_svo_system} are transformed to the systems of ordinary differential equations for Bogoliubov functions~\eqref{eq_bogtransform_1}.
This system can be solved by the standard methods of differential equations, including numerical ones.
After the Bogoliubov functions are calculated, all the averaged values of the field on the output of the crystal for a given input state can be calculated, including number of photons~\eqref{eq_spectra} and quadrature variance~\eqref{eq_variance}.

It should be noted that Bogoliubov transformation~\eqref{eq_bogtransform_1} is valid for any coupling constants and phase mismatches and determines the exact solution of studied system~\eqref{eq_svo_system} that gives a possibility to study the high-gain regime of PDC accompanied by CUpC.

\paragraph{Averaged solution}
\label{par_averaged_solution}

The alternative way to study the CUpC of PDC with nonzero phase-matching is to exploit averaging over the crystal length~$L$ (this approach can be found e.g. in Refs.~\cite{Pe_ina_2015,Chirkin_2002}).
Applying this procedure to the system~\eqref{eq_svo_system} the oscillating terms~$e^{i \Delta z}$ are
\begin{equation}
  \zeta(\Delta) = \dfrac{1}{L} \int_0^L dz \ e^{ i \Delta z} = \mathrm{sinc} \bigg( \dfrac{\Delta L}{2}  \bigg)  e^{ \frac{i \Delta L}{2} },
  \label{eq_averaging_procedure}
\end{equation}
and the wavevector-mismatch is effectively considered by multiplying the initial coupling parameters on the averaged values: $\kappa \rightarrow \kappa \times \zeta(\widetilde\Delta)$, $\eta_i \rightarrow \eta_i \times \zeta(\Delta_i)$, $\eta_s \rightarrow \eta_s \times \zeta(\Delta_s)$.

So the initial system with nonzero phase-matching is replaced by the phase-matched system with reduced coupling constants.
After averaging the equations~\eqref{eq_svo_system} become autonomous and its solutions was obtained for different cases of PDC with CUpC in Refs.~\cite{Smithers_1974,Allevi_2006,Chirkin_2002,Tlyachev_2013,Tlyachev_2014,Arkhipov_2016}.  
For the initial non-autonomous system of equations~\eqref{eq_svo_system} this solution is approximate and differs from the exact one, presented in terms of Bogoliubov transformations~\eqref{eq_bogtransform_1}.

\subsection{Characteristic equation}
\label{subsec_analytical_solution}

For the system of differential equations~\eqref{eq_func_system_1} and \eqref{eq_func_system_2} the analytical solution exists (see Appendix~\ref{appendix_diff_eq_solution}).
In spite of this, its analysis is sophisticated: there are 16 complex Bogoliubov functions which depend on 3 complex parameters $\kappa, \eta_s, \eta_i$ and 3 real ones $\widetilde\Delta, \Delta_s, \Delta_i$.
From the analytical solution for Bogoliubov functions~\eqref{eq_ap_general_form}, \eqref{eq_ap_general_form_of_solution} one can notice that Bogoliubov functions reveal exponential spatial dependence (see Appendix~\ref{appendix_diff_eq_solution})
\begin{equation}
  U(z), V(z) ... \sim e^{\lambda z}.
\end{equation}
Here $\lambda$ are the roots of characteristic equation that has the depressed quartic form (see Appendix~\ref{appendix_diff_eq_solution_characteristic})
\begin{equation}
    \lambda^4 + P\lambda^2 + iQ\lambda + R = 0,
    \label{eq_characteristic_equation}
\end{equation}
where 
\begin{align}
    P &= g_s^2 + g_i^2 + \dfrac{\phi^2}{2} - |\kappa|^2, \\
    Q &= \phi(g_i^2-g_s^2) - |\kappa|^2\dfrac{\Delta_i - \Delta_s}{2}, \\
    R &= \bigg[g_s^2-\dfrac{\phi^2}{4}\bigg]\bigg[g_i^2-\dfrac{\phi^2}{4}\bigg] - \dfrac{|\kappa|^2}{4}(\phi-\Delta_s)(\phi-\Delta_i)
\end{align}
and $g_s^2 = |\eta_s|^2 + \Delta_s^2/4$, $g_i^2 = |\eta_i|^2 + \Delta_i^2/4$, $\phi = \widetilde\Delta - (\Delta_s + \Delta_i)/2$.

So far as the roots of characteristic equation~\eqref{eq_characteristic_equation} are complex, their imaginary parts lead  to the oscillating terms in Bogoliubov functions, while the positive real parts -- to the exponentially increasing contributions.
Thus the parametric amplification exists when at least one of the roots of the characteristic equation has a real positive part. 
At some distance $L$ the exponentially increasing terms predominate over the oscillating contributions and the high-gain regime of PDC with CUpC can be realised (the mean number of photons is larger than~1).  
 
In general case while all the parameters $\kappa, \eta_s, \eta_i, \widetilde\Delta, \Delta_s, \Delta_i$ are independent, the roots of the characteristic equation take sophisticated form.
Nevertheless the nature of the roots for the quartic equation can be obtained from the set of inequalities that involve the discriminant and the parameters $P$, $Q$, $R$ (see Appendix~\ref{appendix_diff_eq_solution_characteristic} and Ref.~\cite{Rees_1922}).
This inequalities define the criteria for parametric amplification and in the next sections are explicitly presented for degenerate, three- and four- mode cases of PDC with CUpC. 

\vspace{0.5cm}

Summarizing this section, the exact solution of CUpC of PDC is presented in terms of Bogoliubov transformation and all the observable values of electromagnetic field and their spatial dynamics along the nonlinear crystal can be expressed in terms of Bogoliubov functions.
The roots analysis of characteristic equation provides the criterion for the parametric amplification and in the next sections we apply the obtained results to the high-gain regime of PDC with CUpC.

\section{Results and discussion: degenerate case}
\label{sec_deg_case}

Let us consider frequency degenerate case when the two interacting modes are present: the PDC mode with the frequency $\omega_{a} = \omega_p/2$ and the wavevector $k_a$; and the CUpC mode with the frequency $\omega_{b} = 3 \omega_p/2$ and the wavevector $k_b$.
In this case $\eta_i = \eta_s$ and $\Delta_i = \Delta_s$ and all the indexes $i$ and $s$ in the Bogoliubov transformation~\eqref{eq_bogtransform_1} can be omitted.

The number of photons in PDC and CUpC modes are
\begin{align}
    \mathcal{N}_{a}(z) &\equiv \langle 0|\hat{a}^{\dagger}(z)\hat{a}(z)|0\rangle = |V(z)|^2 + |Q(z)|^2, \\
     \mathcal{N}_{b}(z) &\equiv \langle 0|\hat{b}^{\dagger}(z)\hat{b}(z)|0\rangle = |L(z)|^2 + |N(z)|^2.
    \label{eq_spectra_deg}
\end{align}

The quadrature variances for PDC $(\Delta X_{a}(\theta_a, z))^2$ and CUpC $(\Delta X_{b}(\theta_b, z))^2$ have the form 
\begin{equation}
(\Delta X_{j}(\theta_j, z))^2 =  1+ 2\mathcal{N}_j(z) + |F_j(z)|\cos(2\theta_j+\varphi_j),
\label{eq_degenerate_variance}
\end{equation}
where index $j = a, b$ corresponds to the PDC and CUpC modes, respectively. 
Here $F_j(z)$ correspond to the correlation functions
\begin{align}
    F_a(z) & \equiv \langle \hat{a}(z) \hat{a}(z) \rangle = U(z)V(z)+W(z)Q(z), \\ 
    F_b(z) & \equiv \langle \hat{b}(z) \hat{b}(z) \rangle = K(z)L(z)+M(z)N(z)
\end{align}
and $\varphi_j = \arg \big[ F_j(z) \big]$.
As stems from~\eqref{eq_degenerate_variance}, the minimal quadrature variances $(\Delta X_{a}^{min})^2$ and $(\Delta X_{b}^{min})^2$ are obtained for the angles $\theta_a = (\pi-\varphi_a)/2$ and $\theta_b = (\pi-\varphi_b)/2$, respectively.

In addition the two-mode squeezing for the collective PDC-CUpC operator can be presented as
\begin{equation}
   \hat{C}(\delta) = \dfrac{ \hat{a}(z) + e^{i\delta}\hat{b}(z) }{\sqrt{2} },
\end{equation}
where $\delta$ is the arbitrary phase. 
Its quadrature variance has the form
\begin{multline}
(\Delta X_{C}(\theta,\delta, z))^2 = 1 + \mathcal{N}_a  +\mathcal{N}_b
 + 2 \mathrm{Re} [G_{ab}(z)e^{i\widetilde{\delta}(z)}] \\ 
 + |F_a(z) + F_b(z)e^{i2\widetilde{\delta}(z)} + 2 F_{ab}(z)e^{i\widetilde{\delta}(z)}| \cos(2\theta+\varphi(z)),
\end{multline}
where $\widetilde\delta(z) = \delta + k_a z - k_b z$ and 
\begin{align}
    F_{ab}(z) &\equiv \langle  \hat{\alpha}(z)\hat{\beta}(z) \rangle  = U(z)L(z) + W(z)N(z), \\
    G_{ab}(z) &\equiv \langle  \hat{\alpha}^{\dagger}(z)\hat{\beta}(z) \rangle = V^*(z)L(z)+Q^*(z)N(z), \\
    \varphi(z) &= \arg \big[F_a(z) + F_b(z)e^{i2\widetilde{\delta}(z)} + 2 F_{ab}(z)e^{i\widetilde{\delta}(z)}\big].
\end{align}
 
The minimal value of the variance $(\Delta X_{C}^{min} (z))^2 = (\Delta X_{C}(\theta_{min}, \delta_{min}, z))^2$ does not have a simple analytical form and should be solved numerically.

\begin{figure*}[t]
    \begin{center}
    \includegraphics[width=0.99\linewidth]{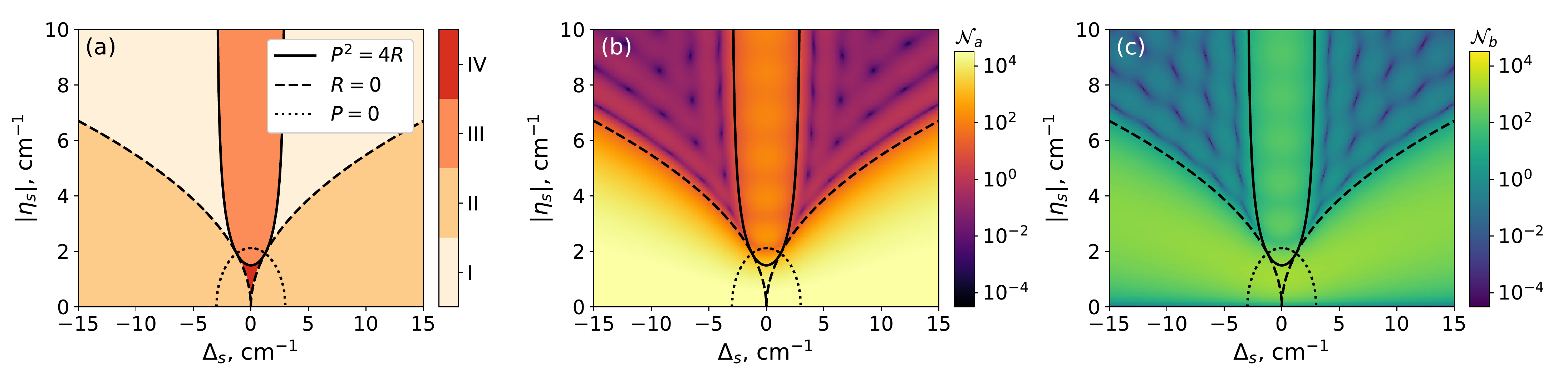} \\
    \includegraphics[width=0.99\linewidth]{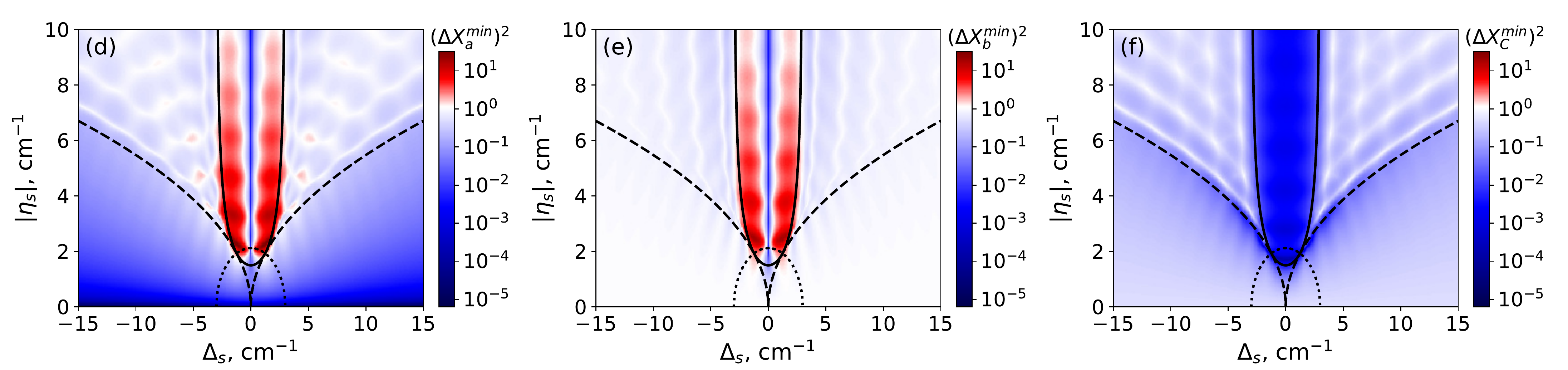}
    \caption{Diagrams $\Delta_s-\eta_s$ for degenerate CUpC of PDC, for the phase-matched PDC case ($\widetilde{\Delta} = 0$, $|\kappa| = 3$~cm$^{-1}$):
    (a) the `phase' diagram (different colors correspond to different regimes of generation);
    (b, c) number of photons $\mathcal{N}_{a}$ (PDC mode) and $\mathcal{N}_{b}$ (CUpC mode), respectively;
    (d, e, f) minimal quadrature variances $(\Delta X^{min}_{a})^2$ (PDC mode),  $(\Delta X^{min}_{b})^2$ (CUpC mode) and $(\Delta X^{min}_{C})^2$ (collective PDC-CUpC), respectively.
    Number of photons and minimal quadrature variances were calculated for the crystal length $L=2$~cm. 
    }
    \label{fig_DC_broad_1}
    \end{center}
\end{figure*}

\begin{figure*}[t]
    \begin{center}
    \includegraphics[width=0.99\linewidth]{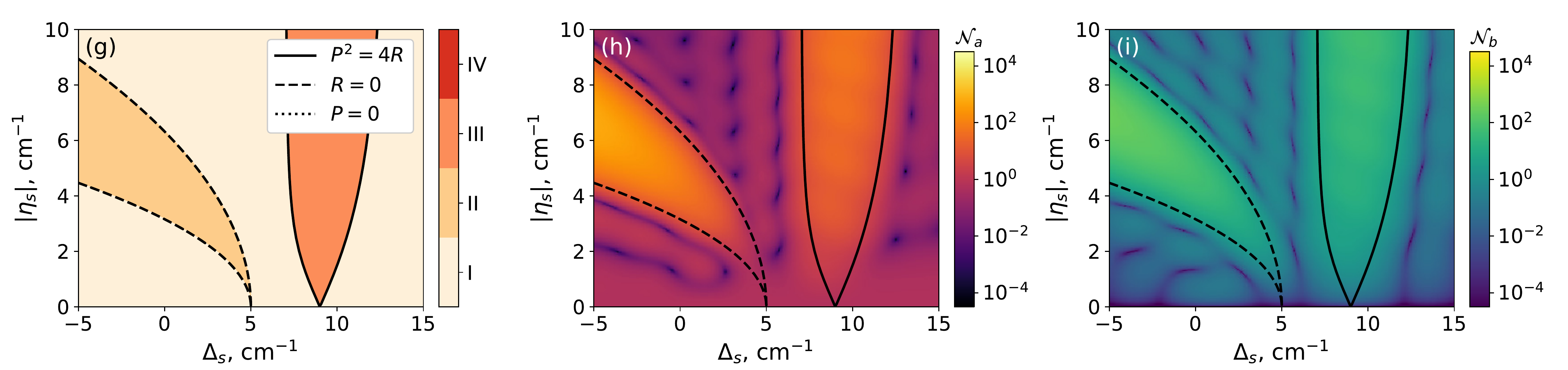} \\
    \includegraphics[width=0.99\linewidth]{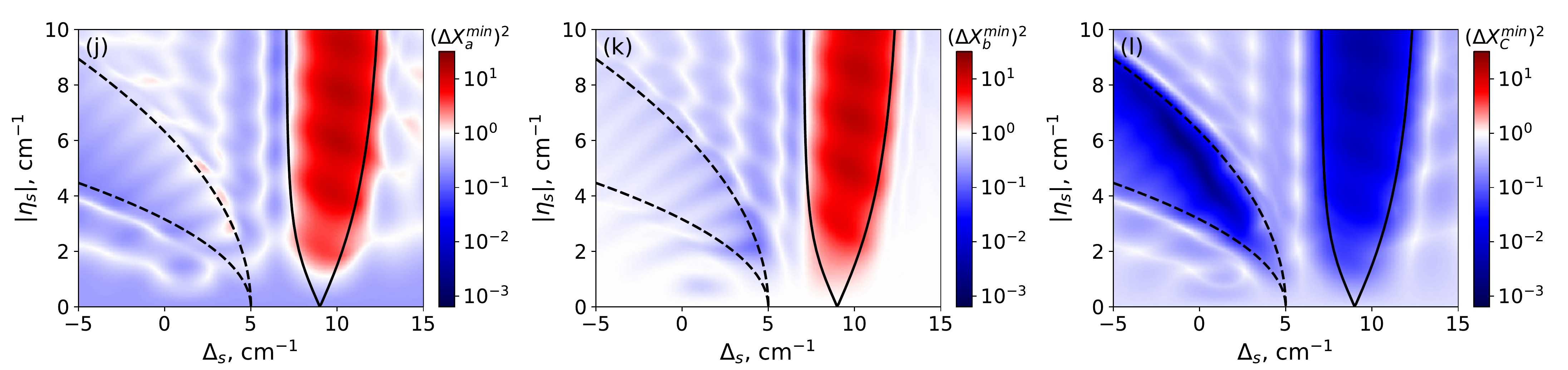}
    \caption{Diagrams $\Delta_s-\eta_s$ for degenerate CUpC of PDC, for the non-phase-matched PDC case ($\widetilde{\Delta} = 10$~cm$^{-1}$, $|\kappa| = 3$~cm$^{-1}$):
    (a) the `phase' diagram (different colors correspond to different regimes of generation);
    (b, c) number of photons $\mathcal{N}_{a}$ (PDC mode) and $\mathcal{N}_{b}$ (CUpC mode), respectively;
    (d, e, f) minimal quadrature variances $(\Delta X^{min}_{a})^2$ (PDC mode),  $(\Delta X^{min}_{b})^2$ (CUpC mode) and $(\Delta X^{min}_{C})^2$ (collective PDC-CUpC), respectively.
    Number of photons and minimal quadrature variances were calculated for crystal length $L=2$~cm.
    }
    \label{fig_DC_broad_2}
    \end{center}
\end{figure*}

\subsection{PDC without CUpC}
\label{subsec_pdc_only}

Before the CUpC properties are considered, we examine our approach for the PDC generation in the absence of CUpC.
In this case we assume that the coupling constants $\eta_s=\eta_i \rightarrow 0$, and the roots of the characteristic equation~\eqref{eq_characteristic_equation} have the form
\begin{equation}
    \lambda_{1,2} = \pm \gamma, \;
    \lambda_{3,4} = \pm i\widetilde\Delta / 2,
    \label{eq_pdc_only_roots}
\end{equation}
where $\gamma = \sqrt{|\kappa|^2 - \widetilde\Delta^2 / 4}$. 
 
By substitution the roots~\eqref{eq_pdc_only_roots} into the solution~\eqref{eq_ap_general_form} and \eqref{eq_ap_general_form_of_solution} with the parameters $\eta_s = \eta_i \rightarrow 0$,  $\Delta_s = \Delta_i \rightarrow 0$ one obtains the nonzero Bogoliubov functions 
\begin{align}
    U(z) &=\Big[\mathrm{cosh}(\gamma z) -\dfrac{i\widetilde{\Delta}}{2\gamma}\mathrm{sinh}(\gamma z)\Big]e^{i\widetilde{\Delta}z/2}, \\
    V(z) &= \dfrac{i\kappa}{\gamma}\mathrm{sinh}(\gamma z)e^{i\widetilde{\Delta}z/2},
    \label{eq_simple_PDC_bogoliubov}
\end{align}
that corresponds to the well known solution for the PDC generation e.g. Refs.~\cite{klyshko1988book,Huttner_1990,Caspani_2010}.

On the output of the crystal with the length $L$ the number of photons for PDC mode has the from 
\begin{equation}
    \mathcal{N}_{a}(L) = |V(L)|^2 = |\kappa|^2 L^2 \bigg[\dfrac{\sinh(\gamma L)}{\gamma L}\bigg]^2.
    \label{eq_num_pdc_only}
\end{equation}
The dimensionless parameter $\Gamma = \gamma L$ is known as the parametric gain for PDC process~\cite{klyshko1988book,Chekhova_2015}.
The case of $\Gamma \gtrsim 1$ is known as high-gain regime, and the mean number of photons in this case $\mathcal{N}_{a} \gg 1$. 
 
One can see that two roots $\lambda_{1,2}$ are real if $|\kappa| > |\widetilde\Delta|/2$ (eq.~\eqref{eq_pdc_only_roots}) and according the eq.~\eqref{eq_num_pdc_only} the number of photons increases exponentially i.e. the parametric amplification exists.
Otherwise all the roots of characteristic equation are imaginary and Bogoliubov functions~\eqref{eq_simple_PDC_bogoliubov} are oscillating.
Thus our statement about determination of parametric amplification by the nature of the roots of characteristic equation (\ref{subsec_analytical_solution}) is valid for PDC generation.

For the phase-matched PDC ($\widetilde{\Delta} = 0$) the minimal variance of the quadrature operator $\hat{X}_{a}(\theta,z) = \hat{a}(z) e^{i\theta} + \hat{a}^{\dagger}(z) e^{-i\theta}$ has form
\begin{equation}
    (\Delta {X}_{a}^{min})^2 = 1 + 2\mathcal{N}_a - 2|UV|  =  e^{-2\Gamma}.
    \label{eq_squeezing_pdc_only}
\end{equation}
So far as the up-conversion process is absent, the CUpC mode remains to be vacuum.

\subsection{Degenerate PDC with CUpC: characteristic equation and roots analysis}
\label{subsec_degcase_roots_analysis}

In this subsection we apply our approach to the analysis of the degenerate PDC with CUpC.
The characteristic equation~\eqref{eq_characteristic_equation} becomes biquadratic ($Q = 0$) with the coefficients
\begin{align}
    P &=  2|\eta_s|^2 + \Delta_s^2 + \dfrac{\widetilde\Delta(\widetilde\Delta - 2\Delta_s)}{2} - |\kappa|^2, \\
    R &= \bigg[ |\eta_s|^2 -\dfrac{\widetilde\Delta(\widetilde\Delta - 2\Delta_s)}{4}\bigg]^2  - \dfrac{|\kappa|^2}{4}(\widetilde\Delta-2\Delta_s)^2 
\end{align}
and the roots are easily obtained 
\begin{equation}
    \lambda_{1,2,3,4}  = \pm \sqrt{\dfrac{-P  \pm \sqrt{P^2-4R}}{2}}.
\end{equation} 

\begin{table}[b!]
\begin{ruledtabular}
\begin{tabular}{ccc}
Area & Condition & Roots \\
\hline
I   &  $P>0$ and $0<R<P^2/4$    & $\lambda_{1,2,3,4} \in \mathbb{I} $ \\  
II  &  $R<0$                    & $\lambda_{1,2} \in \mathbb{R} $, $\lambda_{3,4} \in \mathbb{I} $ \\
III &  $R>P^2/4$                & $\lambda_{1,2,3,4} \in \mathbb{C} $     \\
IV  &  $P<0$ and  $0<R<P^2/4$   & $\lambda_{1,2,3,4} \in \mathbb{R} $     \\ 
V   &  $R=P^2/4$ or $R=0$       & multiple roots \\
\end{tabular}
\caption{Conditions for different cases of roots of characteristic equation for the degenerate CUpC of PDC.  
$\mathbb{R}$ is real number, $\mathbb{I}$ is imaginary value and $\mathbb{C}$ is complex with nonzero real and imaginary part.}
\label{table_deg_case}
\end{ruledtabular}
\end{table}

\begin{figure}[t]
    \begin{center}
    \includegraphics[width=0.99\linewidth]{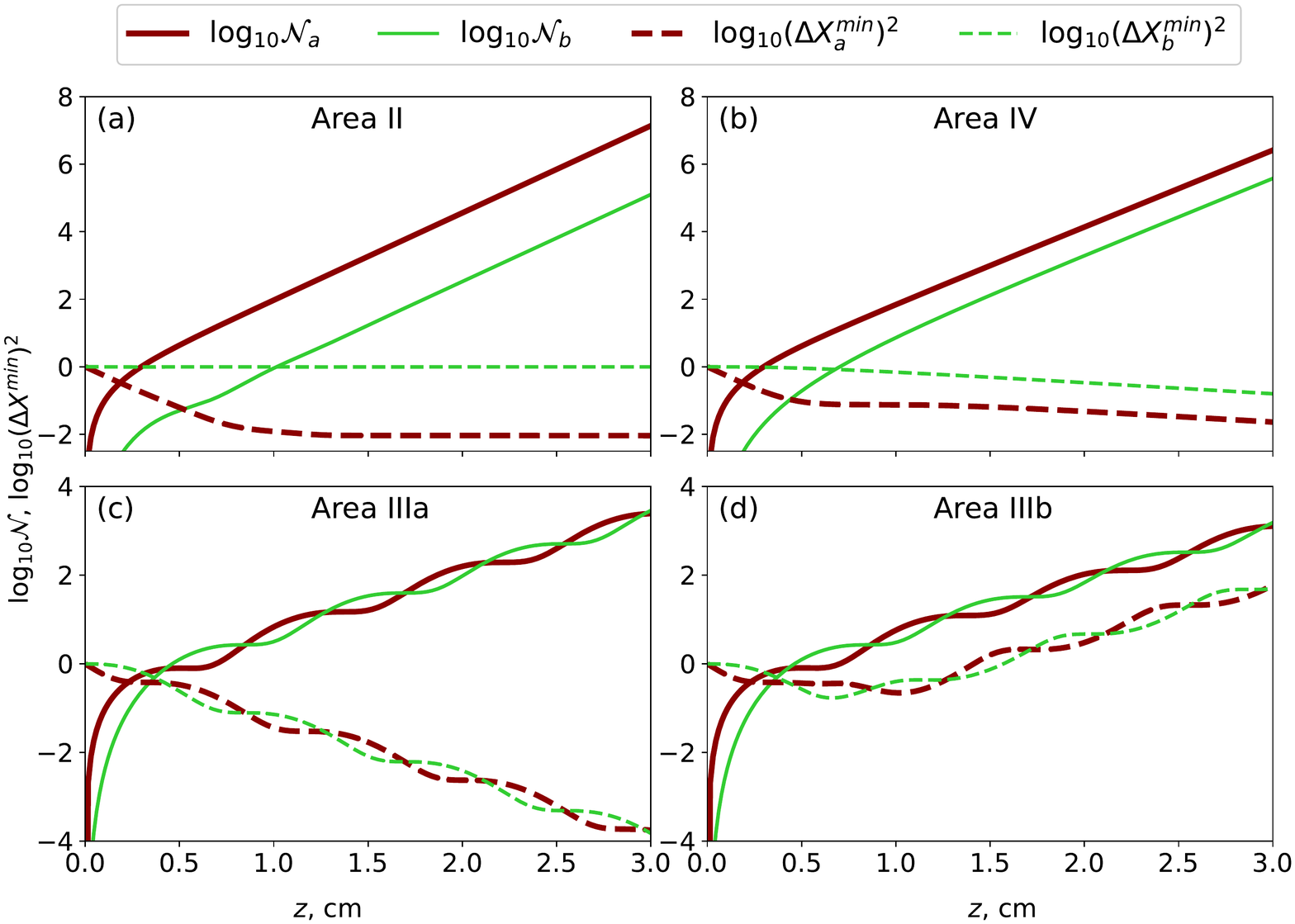}
    \caption{The dependence of number of photons and minimal quadrature variances for PDC and CUpC modes on the nonlinear crystal length for the phase-matched PDC case ($\widetilde{\Delta} = 0$, $|\kappa| = 3$~cm$^{-1}$):\\
    (a) area II ($\Delta_s = 10$~cm$^{-1}$, $\eta_s = 1$~cm$^{-1}$),\\
    (b) area IV ($\Delta_s = 0$~cm$^{-1}$, $\eta_s = 1$~cm$^{-1}$),\\
    (c) area III ($\Delta_s = 0$~cm$^{-1}$, $\eta_s = 4$~cm$^{-1}$),\\
    (d) area III ($\Delta_s = 0.5$~cm$^{-1}$, $\eta_s = 4$~cm$^{-1}$).
  }
    \label{fig_degenerate_case_z_dep}
    \end{center}
\end{figure}

There are 5 possible cases of roots which define the behaviour of solution (Table~\ref{table_deg_case}) that are discussed below.
The description of these cases is accompanied by the diagrams $\Delta_s-\eta_s$ in Figs.~\ref{fig_DC_broad_1}, \ref{fig_DC_broad_2}.
Figures~\ref{fig_DC_broad_1}(a), \ref{fig_DC_broad_2}(a) demonstrate the `phase' diagrams where different color represents different regime of generation.
The analysis is provided for $a=3$~cm$^{-1}$ both for the phase-matched PDC ($\widetilde{\Delta} = 0$, Fig.~\ref{fig_DC_broad_1}) and for the non-phase-matched PDC ($\widetilde{\Delta} = 10$~cm$^{-1}$, Fig.~\ref{fig_DC_broad_2}).
In addition, the number of photons and the minimal variance of quadratures for high-gain regime are also presented in Figs.~\ref{fig_DC_broad_1}, \ref{fig_DC_broad_2} (details are in the caption) for the crystal length $L=2$~cm.
For completeness, the spatial dynamics of photon number and minimal quadrature variance for the PDC and CUpC modes are plotted in Fig.~\ref{fig_degenerate_case_z_dep}.

\paragraph{Area~I}
All the roots $\lambda_{1,2,3,4}$ are imaginary and $\lambda_{3,4} = \lambda^*_{1,2}$.
Generally this is the single regime which corresponds to the absence of the parametric amplification and leads to quasi-periodic solution for Bogoliubov functions.
According Figs.~\ref{fig_DC_broad_1}(b, c), \ref{fig_DC_broad_2}(b, c) inequality $\mathcal{N}_{a, b} < 1$ is valid both for PDC and CUpC modes.

\paragraph{Area~II} 
In this case two roots are real thus the parametric amplification occurs. 
From Figs.~\ref{fig_DC_broad_1}(b, c),~\ref{fig_DC_broad_2}(b, c) one can see that the number of CUpC photons is at least one order of magnitude lower compared to the number of PDC photons.
Increasing of the parameter $\Delta_s$ for the phase-matched PDC generation (Fig.~\ref{fig_DC_broad_1}(b, c)) leads to decreasing of the CUpC efficiency and in this regime the CUpC can be considered as  losses for the PDC radiation.
In details this case is considered below in the subsection~\ref{subsec_DC_losses}.

From Fig.~\ref{fig_degenerate_case_z_dep}(a) one can see that the number of photons both for PDC and CUpC modes increases exponentially all over the nonlinear crystal.
However, the CUpC mode remains unsqueezed and the squeezing of the PDC mode is limited and reaches the constant value during the propagation (for details see subsec.~\ref{subsec_DC_losses}).

For the non-phase-matched PDC generation area~II still exists (Fig.~\ref{fig_DC_broad_2}(b, c)) and parametric amplification for both modes with nonzero wavevector-mismatch can be achieved.

\paragraph{Area~III}
In this regime all the roots are complex numbers with the nonzero real and imaginary parts. 
The parametric amplification inside the crystal is accompanied by the periodic energy transfer between PDC and CUpC modes (Fig.~\ref{fig_degenerate_case_z_dep}(c,d)) and $\mathcal{N}_{a} \approx \mathcal{N}_{b}$ in the whole area~III (Figs.~\ref{fig_DC_broad_1}(b,c), \ref{fig_DC_broad_2}(b,c)).

In the case when both processes are phase-matched ($\widetilde{\Delta} = 0$, $\Delta_s=0$), the PDC and CUpC modes are simultaneously squeezed (Fig.~\ref{fig_DC_broad_1}(d,e)) $(\Delta {X}_{a}^{min})^2 \approx (\Delta {X}_{b}^{min})^2$.
However, a small variation of $\Delta_s$ leads to dramatic change of the single-mode squeezing properties of generated light.
Considering spatial evolution of the minimal single-mode quadrature variances we see that it starts to increase at some point both for PDC and CUpC radiation which is shown in Fig.~\ref{fig_degenerate_case_z_dep}(d).
The qualitative analysis of this effect showed that PDC and CUpC modes can be simultaneously squeezed when $\theta_a=\theta_b+\pi/2$.

Out of the phase-matched PDC ($\widetilde{\Delta} = 0$) both the PDC and CUpC modes are anti-squeezed in the area~III (Fig.~\ref{fig_degenerate_case_z_dep}(d,e)).
However, the two-mode (PDC-CUpC) squeezing for the collective operator $\hat{C}$ (eq.~\eqref{eq_collective_mode}) is always achieved (Figs.~\ref{fig_DC_broad_1}(f), \ref{fig_DC_broad_2}(f)).

\paragraph{Area~IV}
Here all the roots $\lambda_{1,2,3,4}$ are real.
Comparing in Fig.~\ref{fig_degenerate_case_z_dep}(a,b), this regime is similar to the area~II and the most effective generation of CUpC can be obtained.

\paragraph{Area V: multiple roots}
If coupling constants and phase mismatches satisfy conditions $R=0$ or $R=P^2/4$, the roots of the characteristic equation are multiple, and, strictly speaking, the analytical solution in the form presented in Appendix~\ref{appendix_diff_eq_solution} is not valid.
Nevertheless, according to the initial equation~\eqref{eq_ap_start} the solution does not have any singularities and discontinuities in the case of multiple roots and the expressions~\eqref{eq_ap_general_form} and \eqref{eq_ap_general_form_of_solution} can be used in limiting case, taking parameters as close to the curves $R=0$ or $R=P^2/4$ as possible.
The detailed analysis of multiple roots is not considered in this paper.
\vspace{0.5cm}

For degenerate phase-matched PDC with UPC ($\Delta_s = 0$, $\widetilde \Delta=0$) roots have simple form
  \begin{equation}
      \lambda_{1,2,3,4} = \pm \dfrac{|\kappa|}{2} \pm \sqrt{|\kappa|^2/4 -|\eta_s|^2}.
  \end{equation}
If $|\kappa|^2/4 >|\eta_s|^2$, all the roots are real and this corresponds to the area~IV.
When $|\kappa|^2/4 < |\eta_s|^2$ the real part of roots $\lambda_i$ is equal to the $\pm|\kappa|/2$ and is independed on parameter $\eta_s$.
Hence, in area~III we observe some kind of `oscillating plateau' in Figs.~\ref{fig_DC_broad_1}(b, c) for photon numbers and in Fig.~\ref{fig_DC_broad_1}(d,e,f) for one- and two-mode squeezing.
So far as the real part of characteristic equation roots are responsible for the parametric amplification. 
In the high-gain regime $\mathcal{N}_a \approx \mathcal{N}_b \approx \sinh^2 \big( \widetilde\Gamma  \big)$, where $\widetilde\Gamma = |\kappa| L/2 $ is the half from phase-matched PDC generation with the absence of cascade up-conversion.

\subsection{Particular case: strongly phase mismatched CUpC as losses for degenerate PDC}
\label{subsec_DC_losses}

In this subsection we consider one of the important practical cases: phase-matched PDC ($\widetilde\Delta = 0$) with the CUpC with large wavevector-mismatch ($\Delta_s \gg |\kappa|, |\eta_s|$).
For this reason we introduce small parameters $\epsilon_a = |\kappa|/\Delta_s \ll 1$ and $\epsilon_b = |\eta_s|/\Delta_s \ll 1$ and obtain approximation of general solution for this special case. 
The roots of characteristic equation satisfy conditions for the area~II:
\begin{equation}
    \lambda_{1,2} \approx \pm |\kappa|(1 - \epsilon_b^2), \;\;
    \lambda_{3,4} \approx \pm i\Delta_s (1 + \epsilon_b^2).
    \label{eq_roots_losses_case}
\end{equation}

Considering exact solution~\eqref{eq_ap_general_form_of_solution}, approximate number of photons is obtained keeping the first non vanishing term with $\epsilon_b^2$:
\begin{align}
    \mathcal{N}_a & \approx (1 - \epsilon_b^2) \sinh^2 \tilde{\Gamma}, \\ 
    \mathcal{N}_b & \approx \epsilon_b^2 \sinh^2 \tilde{\Gamma},
    \label{eq_app_N}
\end{align}
where $\tilde{\Gamma} = |\kappa| L (1 - \epsilon_b^2)$. 
In this case the CUpC is unefficient compared to PDC and can be assumed as losses for PDC mode. 

As the squeezing properties are sensitive to the losses, we obtain minimal variance of quadrature for the PDC mode
\begin{equation}
     (\Delta {X}_{a}^{min})^2 = (1 - \epsilon_b^2)e^{-2\tilde{\Gamma}} + \epsilon_b^2.
  \label{eq_varQ_approximation}
\end{equation}

It should be noted that for nonlinear crystals the parameters $|\kappa|$ and $|\eta_s|$ linearly depend on the pump amplitude, while the ratio $r = |\eta_s| / |\kappa|$ is determined by the crystal parameters (refractive indexes and effective nonlinear susceptibilities) and does not depend on pump amplitude. 
Thus the PDC squeezing~\eqref{eq_varQ_approximation} as a function of pump power becomes more complicated compared to the PDC generation~\eqref{eq_squeezing_pdc_only}
\begin{equation}
    (\Delta X_a^{min})^2 = (1 - \delta^2 \Gamma^2)e^{-2 \Gamma (1 - \delta^2 \Gamma^2)} +  \delta^2 \Gamma^2,
\end{equation}
where $\delta = r (\Delta_s L)^{-1} $ is small and $\Gamma = |\kappa|L$ is a dimensionless parameter that corresponds to parametric gain of PDC without CUpC~\eqref{eq_num_pdc_only}. 

\begin{figure}[t]
    \begin{center}
    \includegraphics[width=0.99\linewidth]{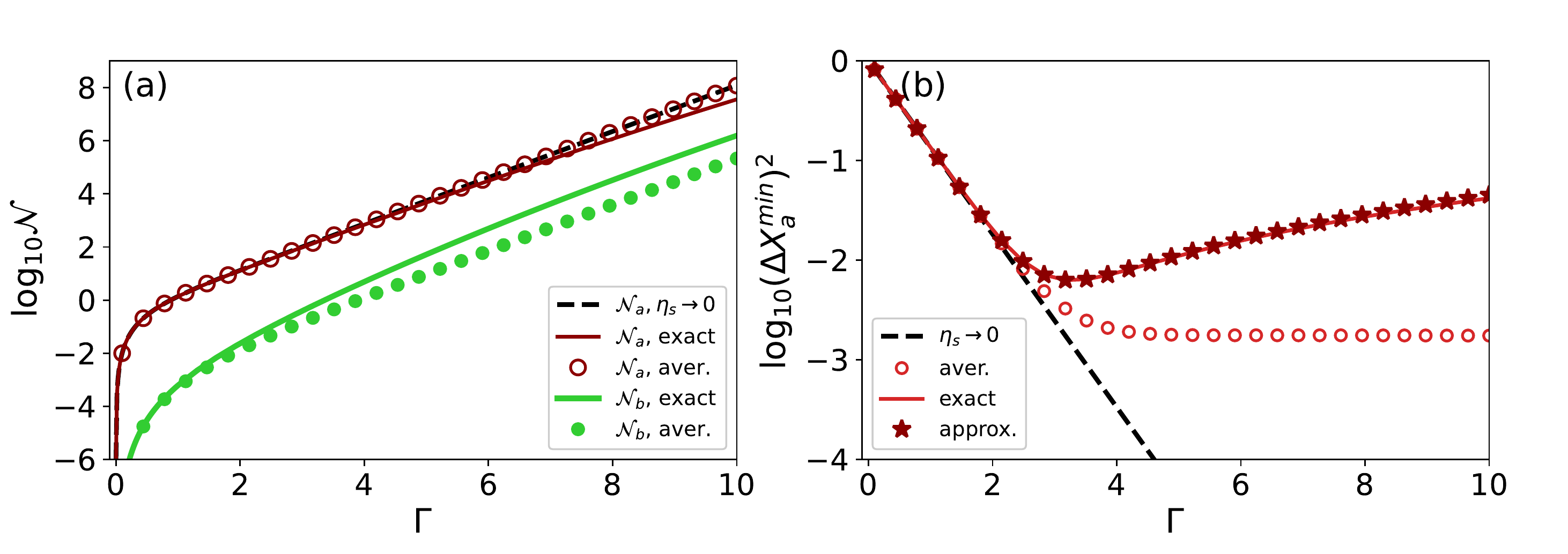}
    \caption{
    (a) Number of photons for PDC and CUpC and (b) minimal quadrature variance for PDC mode as a function of parametric gain $\Gamma$.
    The calculation was obtained with $\Delta_s L = 15\pi$ and $r = |\eta_s|/|\kappa| = 1$ for different models: exact, averaged and for PDC with the absence of CUpC ($\eta_s \rightarrow 0$).
    The approximate expression~\eqref{eq_varQ_approximation} is within good agreement with the exact solution.
    }
    \label{fig_3_squeezing}
    \end{center}
\end{figure}

In Fig.~\ref{fig_3_squeezing}(a) the number of photons for PDC and CUpC modes and in Fig.~\ref{fig_3_squeezing}(b) the minimal variance for the PDC mode $(\Delta {X}_{a}^{min})^2$ are shown as a function of parameter $\Gamma$. 
The calculation was provided with $\Delta_s L = 15\pi$ and $r = 1$ for two models: the exact one (eqs.~\eqref{eq_func_system_1}, \eqref{eq_func_system_2}) and the averaged one obtained using eq.~\eqref{eq_averaging_procedure}.
One can see that the number of photons, calculated by both models, does not significantly differ from each other and for PDC are of the same order as without CUpC~\eqref{eq_num_pdc_only}.

Otherwise, for the minimal quadrature variance different models lead to different results.
In the absence of CUpC the minimal quadrature variance is decreasing with the pump power~\eqref{eq_squeezing_pdc_only}.
The presence of non-phase-matched CUpC leads to the limitation of quadrature squeezing but the calculated dependences calculated with exact and averaged models are completely different: the exact solution provides the local minimum, while the averaged solution results in plateau.

The difference between these two solutions is provided by different regimes of generation: the averaging procedure transfers the solution from area~II into area~IV, that leads to different spatial dynamics of CUpC. 
In addition, the averaged solution is extremely sensitive to the value of $\Delta_s L$: the averaged coupling coefficient $|\zeta(\Delta)|\sim |\mathrm{sinc}(\Delta_s L/2 )|$ (eq.~\eqref{eq_averaging_procedure}) depends periodically on the argument and consequently reaches local maximum when $\Delta_s L = n \pi$ ($n$ is odd).
When $\Delta_s L = m \pi$ ($m$ is even) the coupling averaged coupling coefficient $\zeta(\Delta)=0$ and the CUpC is absent, in opposite to the exact solution that always gives the non-zero CUpC and does not strongly depend on $\Delta_s L$.

A few words should be devoted to the necessity of taking into consideration CUpC when squeezed states via PDC are generated.
One can notice that the maximal squeezing of the PDC mode is obtained when $\mathcal{N}_b \approx 1$ and the condition $\mathcal{N}_b < 1$ can be treated as the criterion for optimal generation of squeezed PDC states. 

\section{Results and discussion: non-degenerate regime}
\label{sec_nondeg_case}

In the previous section the degenerate regime of PDC with CUpC was analysed.
In this section we consider more complicated cases of non degenerate regime of generation: three- and four-mode interaction.

\subsection{Three-mode interaction}

Let us consider the non-degenerate PDC generation when only one PDC mode (e.g. signal) is up-converted.
This regime have been previously studied within simultaneous phase-matching for PDC with CUpC~\cite{Tang_1969,Mishkin_1969,Smithers_1974}.
It is shown that all three waves are parametric amplified if $|\kappa| > |\eta_s|$ and are oscillating if $|\kappa| < |\eta_s|$.
Below, with the use of our approach, we extend this criterion on the non-phase-matched case.

So far as only signal PDC mode is up-converted, we set $\eta_i \rightarrow 0$ and $\Delta_i \rightarrow 0$.
In this case one of the roots of the initial quartic equation~\eqref{eq_characteristic_equation} is always imaginary $\lambda_4 = i\phi/2$ and the equation is reduced to the cubic one
\begin{equation}
      \bigg[\lambda^2-|\kappa|^2+\dfrac{\phi^2}{4} + g_s^2\bigg]\bigg[\lambda + \dfrac{i\phi}{2}\bigg] - i\phi g_s^2 + \dfrac{i\Delta_s}{2}|\kappa|^2=0.
\end{equation}

For the cubic polynomial the discriminant has the form $D_3 = -4P_3^3 - 27Q_3^2$, where 
\begin{align}
  P_3 &= g_s^2-|\kappa|^2+\dfrac{\phi^2}{3} , \\
  Q_3 &= \dfrac{\Delta_s}{2}|\kappa|^2 - \dfrac{2\phi}{3}\bigg[g_s^2 + \dfrac{|\kappa|^2}{2} - \dfrac{\phi^2}{9}\bigg],
\end{align}
and $g_s^2 = |\eta_s|^2 + \Delta_s^2/4$, $\phi = \widetilde{\Delta} - \Delta_s/2$.

Thus for the three-mode interaction three regimes are realized:
\begin{enumerate}
  \item $D_3<0$ oscillating solution for Bogoliubov functions (all the roots are imaginary).
  \item $D_3>0$ -- parametric amplification (it should be noted that according the Vieta's formulas $\mathrm{Re}[\lambda_1 + \lambda_2 + \lambda_3] = 0$ and at least one of the root is always real and positive).
  \item $D_3=0$ -- multiple roots.
\end{enumerate}

\begin{figure}[h!]
      \includegraphics[width=0.99\linewidth]{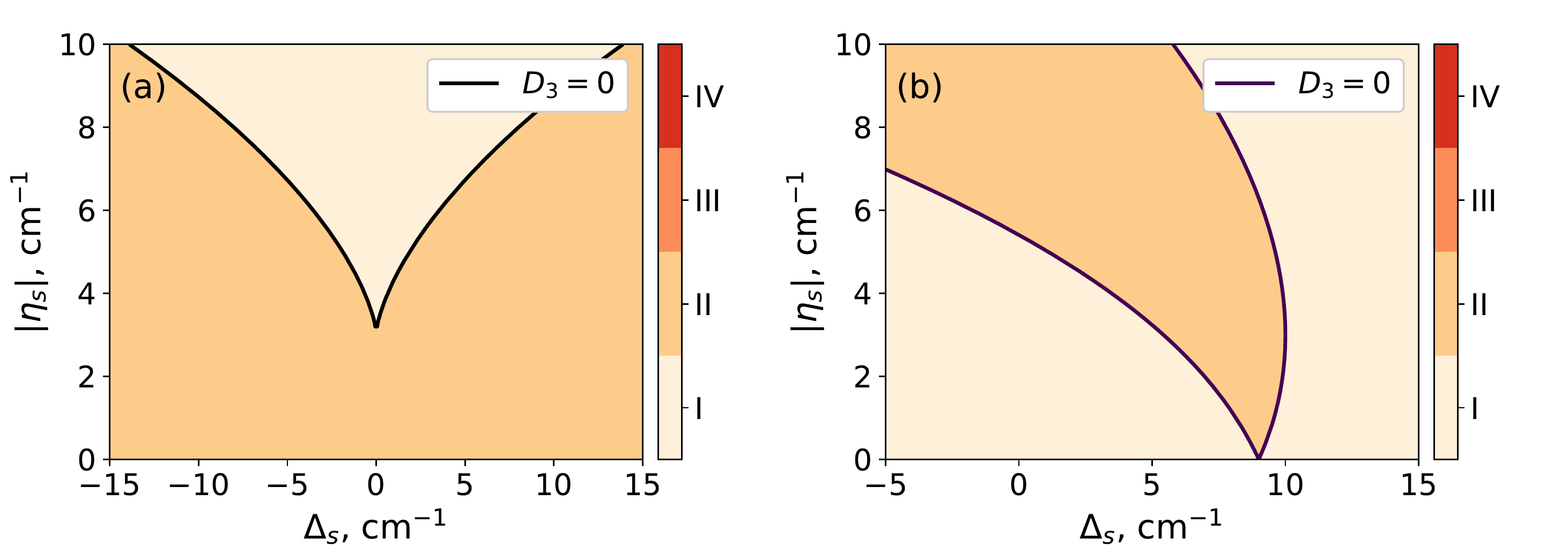}
      \caption{ `Phase' diagrams $\eta_s - \Delta_s$ for three-mode interaction for (a) phase-matched PDC $\widetilde{\Delta} = 0$ and (b) non-phase-matched PDC $\widetilde{\Delta} = 10$~cm$^{-1}$. Area~I: oscillating solution for Bogoliubov functions; area~II  parametric amplification exists.
      }
    \label{fig_three_modes}
\end{figure}

Figure~\ref{fig_three_modes}(a) demonstrates the `phase' diagram $\eta_s - \Delta_s$ for the phase-matched PDC generation $\widetilde{\Delta} = 0$~cm$^{-1}$ for $\kappa = 3$~cm$^{-1}$.
The `phase' diagram  $\eta_s - \Delta_s$  for the non-phase-matched PDC ($\widetilde{\Delta} = 10$~cm$^{-1}$ and $\kappa = 3$~cm$^{-1}$) is shown in Fig.~\ref{fig_three_modes}(b).
One can see, that the parametric amplification exists even if both the processes PDC and CUpC are non-phase-matched.

\subsection{Four-mode interaction: general analysis}

Generally, the CUpC is present for both the signal and idler waves and all the parameters $\kappa, \eta_s, \eta_i, \widetilde\Delta, \Delta_s, \Delta_i$ are independent.
According to Ref.~\cite{Rees_1922} the nature of the roots of characteristic equation~\eqref{eq_characteristic_equation} is determined by the discriminant
\begin{multline}
    D = 256 R^3 - 128 P^2 R^2 + 144 PQ^2R \\ - 27Q^4 + 16P^4R - 4P^3Q^2, 
    \label{eq_discriminant}
\end{multline}
and in TABLE~\ref{table_nondeg_case} the different regimes are shown.
In analogous way to the degenerate case (subsection~\ref{subsec_degcase_roots_analysis}) the parametric amplification is realized in the areas II and III, while the oscillating solution for the Bogoliubov functions exists in the area~I.
\begin{table}[h!]
\begin{ruledtabular}
\begin{tabular}{ccc}
Area & Condition & Roots \\
\hline
I   &  $D>0$ and $P>0$  and $R<P^2/4$ & $\lambda_{1,2,3,4} \in \mathbb{I}$ \\   
II  &  $D<0$                          & $\lambda_{1,2} \in \mathbb{R} $, $\lambda_{3,4} \in \mathbb{I} $ \\
III &  ($D>0$ and $P>0$ and $R>P^2/4$) &  $\lambda_{1,2,3,4} \in \mathbb{C} $   \\
    &  or ($D>0 , P \leq 0$) &  \\
V   &  $D=0$ & multiple roots \\
\end{tabular}
\caption{Conditions for different natures of roots of characteristic equation for the non-degenerate four-mode CUpC of PDC.  
$\mathbb{R}$ is real number, $\mathbb{I}$ is imaginary value and $\mathbb{C}$ is complex with nonzero real and imaginary part.}
\label{table_nondeg_case}
\end{ruledtabular}
\end{table}

\subsection{Four-mode interaction: cascaded phase-matching}

In this subsections we confine the discussion to some particular cases of four-mode interaction, when the cascaded phase-matching can be achieved.
For PDC with CUpC the cascaded wavevector mismatches are introduced
\begin{align}
  \Phi_s &= \widetilde\Delta -\Delta_s = 2 k_p - k_{ai} - k_{bs}, \\
  \Phi_i &= \widetilde\Delta -\Delta_i = 2 k_p - k_{as} - k_{bi}, \\
  \Phi_{si} &= \widetilde\Delta -\Delta_s - \Delta_i = 3 k_p - k_{bs} - k_{bi}.
  \label{eq_cascaded_pm}
\end{align}

It should be noted that cascaded phase-matching can be realised when all the processes are simultaneously non-phase-matched: $\widetilde{\Delta} \neq 0$, $\Delta_s \neq 0$ and $\Delta_i \neq 0$.

\begin{figure}[h!]
      \includegraphics[width=0.89\linewidth]{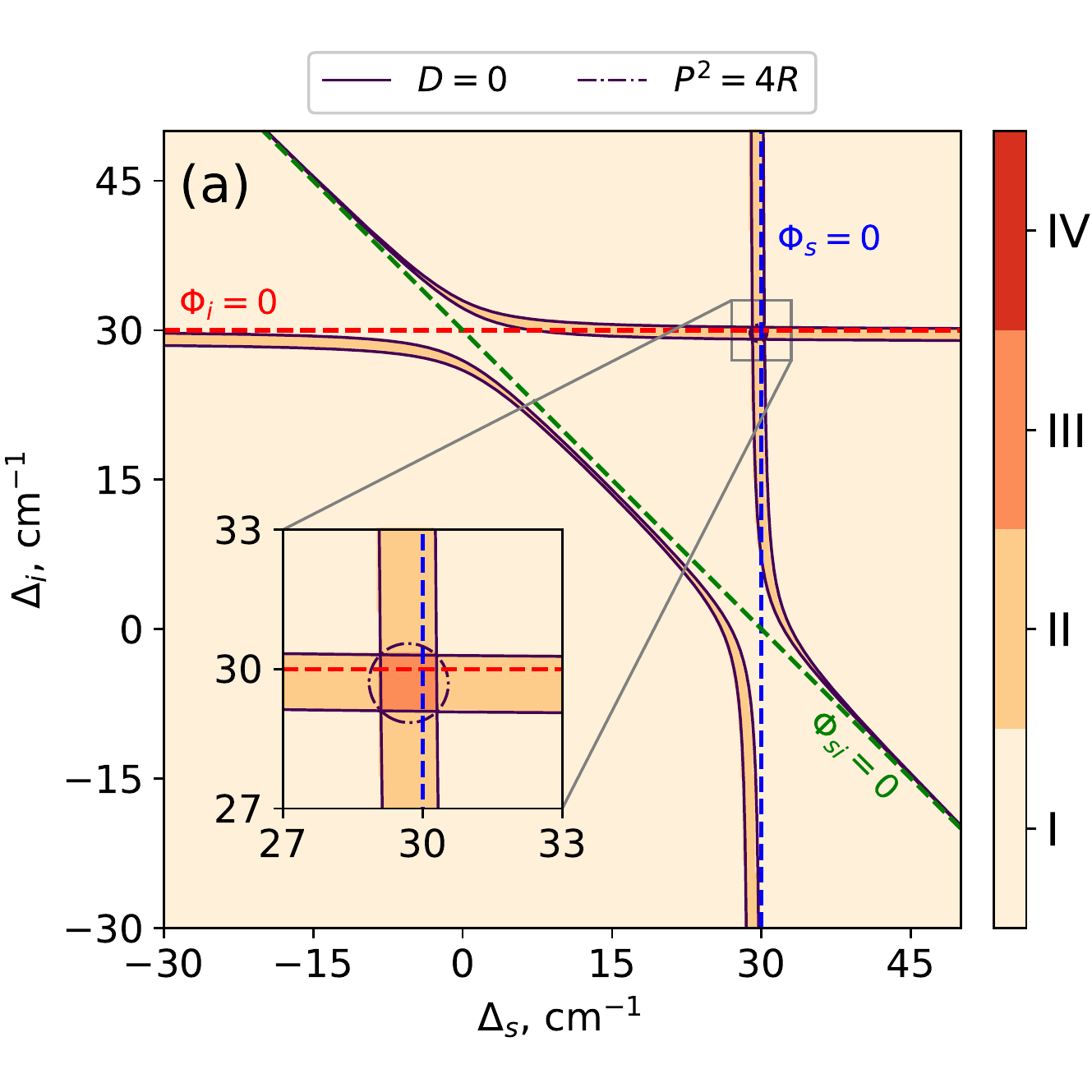}\\
      \includegraphics[width=0.99\linewidth]{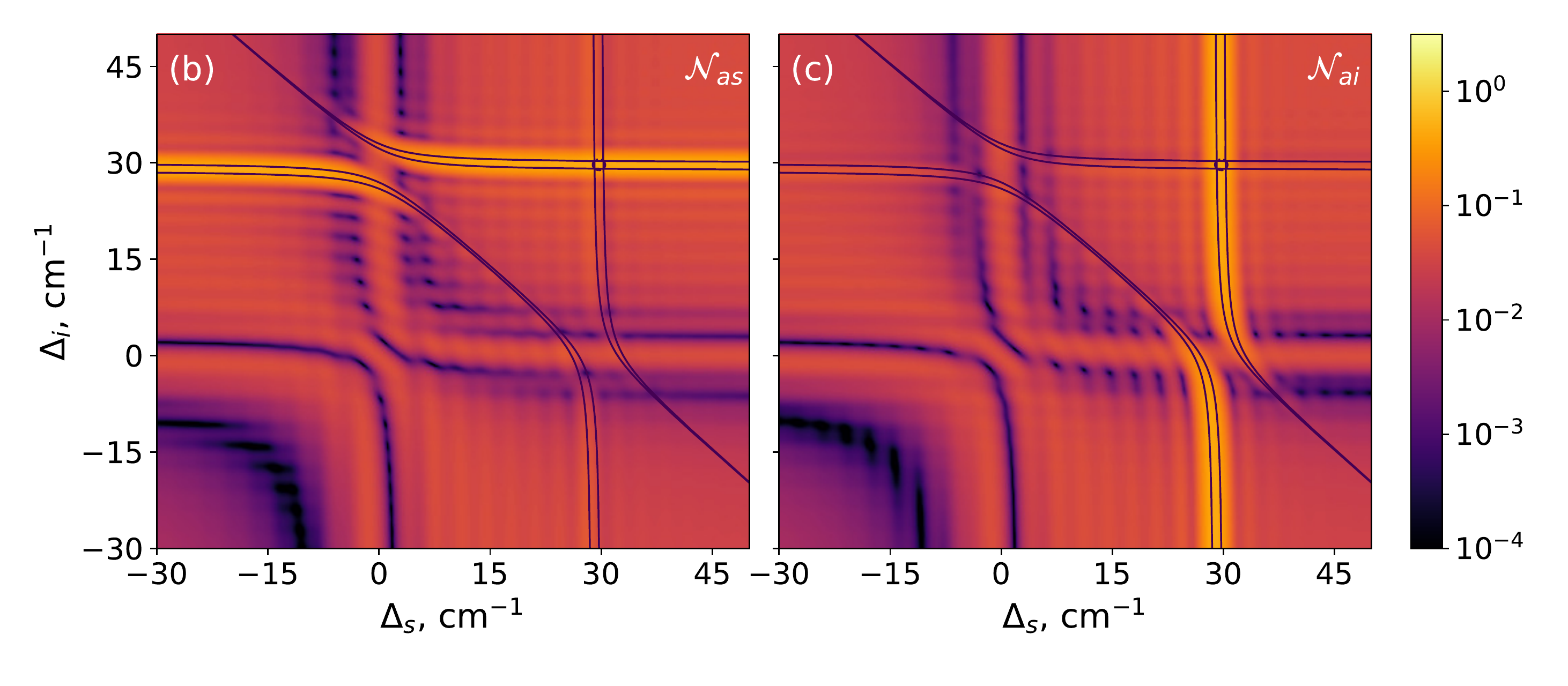}\\
      \includegraphics[width=0.99\linewidth]{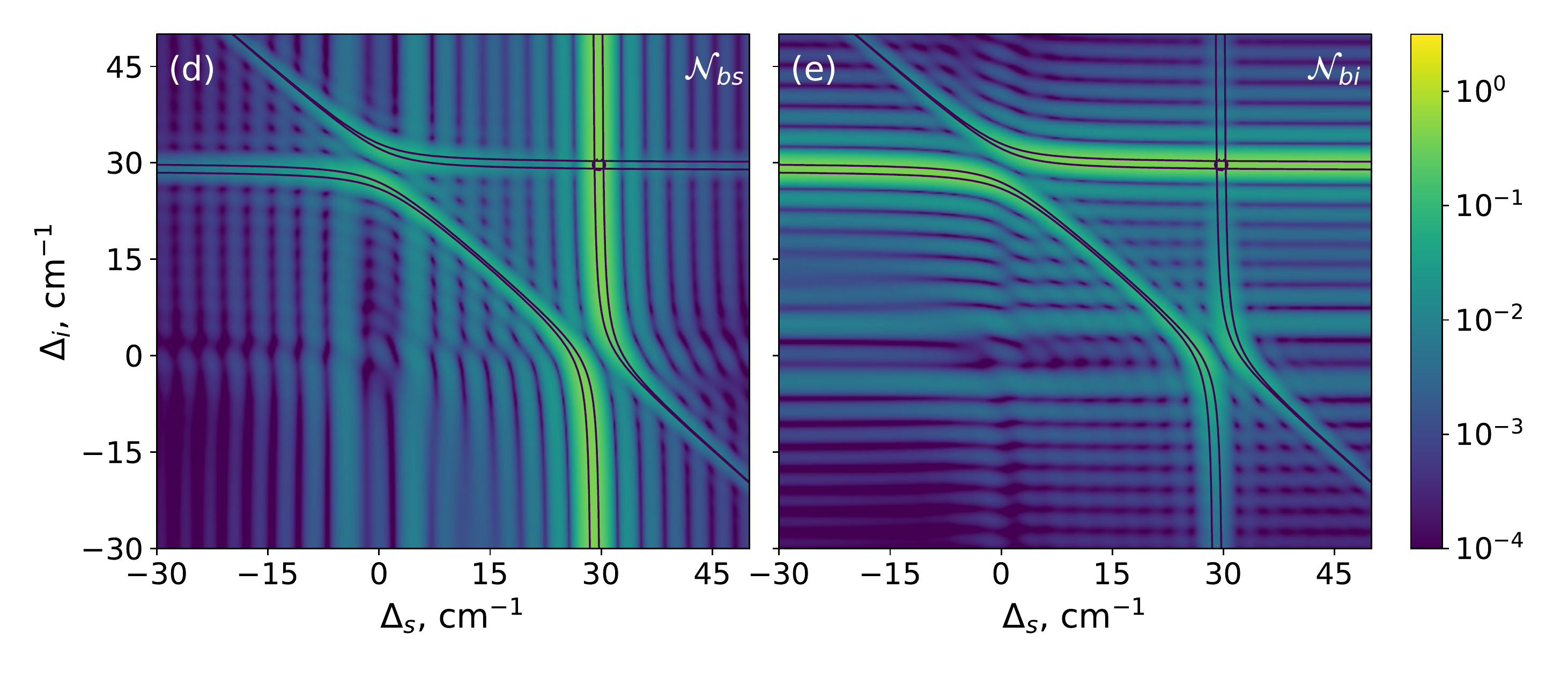}
    \caption{Diagrams $\Delta_i-\Delta_s$ for four-mode CUpC with non-phase-matched PDC $\widetilde{\Delta} = 0$. The coupling coefficients are $\kappa=\eta_s=\eta_i=3$~cm$^{-1}$.
    (a) the `phase' diagram (different colors correspond to different areas from TABLE~\ref{table_nondeg_case}). The dashed lines correspond to different cascaded phase-matching.
    (b, c, d, e) are the numbers of photons in PDC signal, PDC idler, CUpC signal and CUpC idler modes, respectively, calculated for crystal length $L=2$~cm.
    }
    \label{fig_four_modes}
\end{figure}

Figure~\ref{fig_four_modes}(a) shows the `phase'-diagrams `$\Delta_s - \Delta_i$' calculated for non-phase-matched PDC with $\widetilde{\Delta} = 30$~cm$^{-1}$ and coupling parameters $\kappa=\eta_s=\eta_i=3$~cm$^{-1}$.
In spite of the fact that PDC is non-phase-matched, the parametric amplification exists (areas~II and areas~III) and the high-gain regime can be achieved. 
Strictly speaking, the areas~II and III are defined by the conditions from TABLE~\ref{table_nondeg_case}, however their location in the `phase'-diagrams is close to the cascaded phase-matching conditions (eqs.~\eqref{eq_cascaded_pm}): the areas~II corresponds to the cascaded phase-matching conditions $\Phi_s\approx0$ and $\Phi_i\approx0$.
In the `phase' diagram on the concurrence of lines $\Phi_s\approx0$ and $\Phi_i\approx0$ the area~III appears.

In Figure~\ref{fig_four_modes}(b-d) the number of photons for different modes are present for high-gain regime.
One can notice that in the case of $\Phi_i \approx 0$ (red dashed line in Figure~\ref{fig_four_modes}(a)) the amplification exists for signal PDC mode and idler CUpC mode $\mathcal{N}_{as} \approx \mathcal{N}_{bi}$ in Figures~\ref{fig_four_modes}(a), (e).
In the case of $\Phi_s \approx 0$ (blue dashed line in Figure~\ref{fig_four_modes}(a)) the situation is opposite: amplification exists for idler PDC mode and signal CUpC mode (Figures~\ref{fig_four_modes}(c), (d)). 
In the area~III, near the intersection of red and blue dashed lines in Figure~\ref{fig_four_modes}(a), the intensity of all the modes are $\mathcal{N}_{as} \approx \mathcal{N}_{bs} \approx \mathcal{N}_{ai} \approx \mathcal{N}_{bi}$.
For the cascaded phase-matching  $\Phi_{si}=0$ one can notice that the number of photons in CUpC modes is higher compared to the PDC ones.

\vspace{0.5cm}
Summing up, in this section the criteria for the parametric amplification in three- and four-mode generation are obtained.
The parametric amplification is present even if PDC and CUpC are separately non-phase-matched and the cascaded phase-matching can be realized.

\section{Conclusions}

In summary, the exact solution in terms of Bogoliubov transformation for PDC with CUpC with non-zero wavevector-mismatches is presented.
The simple relations, based on roots analysis of characteristic equation, for oscillating and parametric amplification regimes for CUpC of PDC are obtained for degenerate, three- and four-mode generation.
We demonstrate that the high-gain regime of PDC with CUpC is determined not only by phase-matching conditions for each separate process, but also by the cascaded phase-matching conditions.

For the degenerate PDC generation we show that the CUpC can be assumed as dynamical losses for the PDC radiation.
The presence of CUpC leads to the limitation of minimal quadrature dispersion of PDC mode that can be crucial for practical generation of squeezed light via PDC.

So far as the solution is obtained in Heisenberg picture and has the form of Bogoliubov transformation, it is valid for any initial state of light.
Here we confine ourself to the vacuum input state and present the analysis of mean number of photons and quadrature dispersion squeezing of interacted modes. 
However, from the general point of view the studied system corresponds to the class of `Gaussian unitaries'~\cite{Weedbrook_2012} and all the methods for such systems can be applied for CUpC of PDC.

The results obtained in this paper can be used for the noise analysis in quantum frequency converters and for the development of new types of entangled and squeezed visible and UV light sources based on nonlinear crystals, periodically and a-periodically poled nonlinear crystals and nonlinear waveguides. 

\begin{acknowledgments}
Authors thank T.V. Murzina for helpful discussions and supporting.
A.V.R acknowledge the financial support of Theoretical Physics and Mathematics Advancement Foundation `BASIS' (20-2-9-8-1).
\end{acknowledgments}

\appendix 

\section{Bogoliubov transformation}
\label{appendix_bogoliubov}

In general the Bogoliubov transformation can be written in the form~\cite{Weedbrook_2012}
\begin{equation}
\begin{bmatrix} 
    \mathbf{\hat{b}} \\
    \mathbf{\hat{b}}^{\dagger} 
 \end{bmatrix}
 =
 \begin{bmatrix}  
        \mathcal{A}         &   \mathcal{B} \\ 
        \mathcal{B}^{\ast} &    \mathcal{A}^{\ast} 
 \end{bmatrix}
\begin{bmatrix} 
    \mathbf{\hat{a}} \\
    \mathbf{\hat{a}}^{\dagger} 
 \end{bmatrix}
 + 
 \begin{bmatrix} 
    {d} \\
    {d}^{\dagger} 
 \end{bmatrix}.
\end{equation}
For the system, studied in this paper
\begin{align}\begin{split}
\mathbf{\hat{b}} &= [\hat{\alpha}_s(z), \hat{\alpha}_i(z), \hat{\beta}_s(z), \hat{\beta}_i(z)]^T, \\
\mathbf{\hat{b}}^{\dagger} &= [\hat{\alpha}_s^\dagger(z), \hat{\alpha}_i^\dagger(z), \hat{\beta}_s^\dagger(z), \hat{\beta}_i^\dagger(z)]^T,\\
\mathbf{\hat{a}} &= [\hat{\alpha}_s(0), \hat{\alpha}_i(0), \hat{\beta}_s(0), \hat{\beta}_i(0)]^T, \\
\mathbf{\hat{a}}^{\dagger}  &= [\hat{\alpha}_s^\dagger(0), \hat{\alpha}_i^\dagger(0), \hat{\beta}_s^\dagger(0), \hat{\beta}_i^\dagger(0)]^T,
\end{split}\end{align}
and $d=[0,0,0,0]^T$, ${d}^{\dagger}=[0,0,0,0]^T$.

According to the Heisenberg equation~\eqref{eq_svo_system} the system is closed under operators $\hat{\alpha}_s(z)$, $\hat{\alpha}_i^{\dagger}$, $\hat{\beta}_s(z)$, $\hat{\beta}^\dagger_i(z)$, consequently the matrices $\mathcal{A}$ and $\mathcal{B}$ have the form
\begin{equation}
\mathcal{A} = 
\begin{bmatrix}  
  U_s(z)  & 0       & W_s(z)  & 0\\
  0       & U_i(z)  & 0       & W_i(z)\\
  K_s(z)  & 0       & M_s(z)  & 0 \\
  0       & K_i(z)  & 0       & M_i(z)
 \end{bmatrix}
\end{equation}
\begin{equation}
 \mathcal{B}  = 
\begin{bmatrix}  
0       & V_s(z)  & 0       & Q_s(z) \\
V_i(z)  & 0       & Q_i(z)  & 0 \\
0       & L_s(z)  & 0       & N_s(z) \\
L_i(z)  & 0       & N_i(z)  & 0  \\
 \end{bmatrix}
\end{equation}
So far as Bogoliubov transformation is canonical (operators $\hat{\alpha}_s(z), \hat{\alpha}_i(z), \hat{\beta}_s(z), \hat{\beta}_i(z)$ are bosonic with the commutation relations~\eqref{eq_commutation}) the following conditions should be satisfied for any $z$
\begin{equation}
\mathcal{A} \mathcal{A} ^{\dagger} - \mathcal{B}\mathcal{B}^{\dagger} = \mathbb{I}_4, \; \mathcal{A} \mathcal{B}^{T} = (\mathcal{A} \mathcal{B}^{T})^{T}.
\end{equation}
The equalities for Bogoliubov functions 
\begin{align}
    |U_{s}|^2 + |W_{s}|^2 - |V_{s}|^2 - |Q_{s}|^2 &= 1, \\
    |K_{s}|^2 + |M_{s}|^2 - |L_{s}|^2 - |N_{s}|^2 &= 1,
\end{align}
\begin{align}
    U_{s}^* K_{s} + W_{s}^* M_{s} &= V_{s}^* L_{s} + Q_{s}^* N_{s},\\
    U_{s} V_{i} + W_{s} Q_{i} &= U_{i} V_{s} + W_{i} Q_{s},\\
    K_{s} L_{i} + M_{s} N_{i} &= K_{i} L_{s} + M_{i} N_{s},\\
    U_{s} L_{i} + W_{s} N_{i} &= K_{i} V_{s} + M_{i} Q_{s}.
\end{align}
The same conditions are valid for replaced indexes $i \leftrightarrow s$.

\section{Analytical solution of differential systems}
\label{appendix_diff_eq_solution}

Let us consider the differential system 
\begin{equation}
    \begin{cases}
    \frac{d}{dz} Y_1 =i a e^{i\Delta_1z}Y_2+ i b^* e^{i\Delta_{2}z} Y_3,
    \\
    \frac{d}{dz} Y_2 =-ia^* e^{-i\Delta_1z}Y_1 - i c e^{-i\Delta_{3}z} Y_4,
    \\
    \frac{d}{dz} Y_3 =i b e^{-i\Delta_{2}z}Y_1,
    \\
    \frac{d}{dz} Y_4 =-i c^* e^{i\Delta_{3}z}Y_2.
    \end{cases}
    \label{eq_ap_start}
\end{equation}

By excluding $Y_3$ and $Y_4$ and introduction the new functions $\bar{Y}_1= Y_1e^{-i \Delta_1 z/2 + i(\Delta_3 - \Delta_2)z/4}$, $\bar{Y}_2= Y_2e^{ i\Delta_1 z/2 + i(\Delta_3 - \Delta_2)z/4}$ one can obtain autonomous system
\begin{equation}
    \begin{cases}
    \bigg[\bigg(\dfrac{d}{dz} + \dfrac{i\phi}{2}\bigg)^2 + g_b^2\bigg] \bar{Y}_1 =ia\bigg[\dfrac{d}{dz} + i\dfrac{\phi - \Delta_2}{2}\bigg]\bar{Y}_2,
    \\
    \bigg[\bigg(\dfrac{d}{dz} - \dfrac{i\phi}{2}\bigg)^2 + g_c^2 \bigg] \bar{Y}_2 =-ia^{\ast}\bigg[\dfrac{d}{dz} - i\dfrac{\phi - \Delta_3}{2}\bigg]\bar{Y}_1,
    \end{cases}
\end{equation}
where $g_b^2 = |b|^2 + \Delta_2^2/4$, $g_c^2 = |c|^2 + \Delta_3^2/4$, $\phi = \Delta_1 - (\Delta_2 + \Delta_3)/2$.
Finally we get single differential equation of the fourth degree for the $\bar{Y}_1$:  
\begin{equation}
    \bigg[\dfrac{d^4}{dz^4} + P\dfrac{d^2}{dz^2} + iQ\dfrac{d}{dz} + R\bigg]\bar{Y}_1 =0,
\end{equation}
where coefficients are given:
\begin{align}
    P &= g_b^2 + g_c^2 + \dfrac{\phi^2}{2} - |a|^2,\\
    Q &= \phi(g_c^2-g_b^2) - |a|^2\dfrac{\Delta_3 - \Delta_2}{2},\\
    R &= \bigg[g_b^2-\dfrac{\phi^2}{4}\bigg]\bigg[g_c^2-\dfrac{\phi^2}{4}\bigg] - \dfrac{|a|^2}{4}(\phi-\Delta_3)(\phi-\Delta_2).
\end{align}
The characteristic equation for this equation has the form
\begin{equation}
    \lambda^4 + P\lambda^2 + iQ\lambda + R = 0,
    \label{eq_ap_depressed_form}
\end{equation}

In the case of nonzero discriminant of the equation~\eqref{eq_ap_depressed_form} the equation has distinct roots and the function $Y_1$ takes following form
\begin{equation}
    Y_1(z) = \sum_k \widetilde{C}_k e^{ \alpha_k z},
    \label{eq_ap_general_form}
\end{equation}
where $\alpha_k \equiv \lambda_k + i (2\Delta_1 + \Delta_2 - \Delta_3)/4$ and coefficients $\widetilde{C}_k$ are determined by the initial conditions.

The functions $Y_2(z)$, $Y_3(z)$, $Y_4(z)$ can be obtained from $Y_1(z)$ as
\begin{align}\begin{split}
  Y_3(z) &= i b \int_0^z dz^\prime \ e^{-i \Delta_2 z^\prime} Y_1(z^\prime), \\
  Y_2(z) &= \dfrac{e^{-i\Delta_1 z}}{ia}\dfrac{\partial Y_1(z)}{\partial z} - \dfrac{b^* e^{i(\Delta_2-\Delta_1) z}}{a} Y_3(z), \\
  Y_4(z) &= -i c^* \int_0^z dz^\prime \ e^{i \Delta_3 z^\prime} Y_2(z^\prime).
  \label{eq_ap_general_form_of_solution}
\end{split}\end{align}

\subsection{Analytical solution for Bogoliubov functions $U(z),\ V(z),\ K(z),\ L(z)$ }

It can be noticed that differential system~\eqref{eq_func_system_1} for the functions $U(z),\ V(z),\ K(z),\ L(z)$ has the form~\eqref{eq_ap_start} with the initial conditions $Y_1(z)=1$, $Y_2(z)=0$, $Y_3(z)=0$, $Y_4(z)=0$.

According the~\eqref{eq_ap_general_form} the function has the form
\begin{equation}
    Y_1(z) = \sum_k C_k e^{ \alpha_k z},
    \label{eq_ap_first_y1}
\end{equation}
where $\alpha_k \equiv \lambda_k + i \Delta_1/2 - i(\Delta_3 - \Delta_2)/4$.

By substitution the solution~\eqref{eq_ap_first_y1} into the initial conditions
\begin{align}\begin{split}
    & Y_1(0)=1, \ \dfrac{\partial Y_1}{\partial z}(0)= 0, \
    \dfrac{\partial^2 Y_1}{\partial z^2}(0)= |a|^2-|b|^2, \\
    &\dfrac{\partial^3 Y_1}{\partial z^3}(0)= i(\Delta_1 |a|^2 - \Delta_2 |b|^2),
\end{split}\end{align}
the coefficients $C_k$ are determined by the equation
\begin{equation}
\begin{pmatrix}  1 & 1 & 1& 1 \\ \alpha_1& \alpha_2 & \alpha_3 &\alpha_4 \\ \alpha^2_1& \alpha^2_2 & \alpha^2_3 &\alpha^2_4 \\ \alpha^3_1& \alpha^3_2 & \alpha^3_3 &\alpha^3_4 \end{pmatrix} \begin{pmatrix} C_1 \\ C_2 \\ C_3 \\ C_4 \end{pmatrix}=  \begin{pmatrix} 1 \\ 0 \\ |a|^2-|b|^2 \\  i(\Delta_1|a|^2-\Delta_2|b|^2) \end{pmatrix}
\end{equation}

The explicit form of the functions is
\begin{multline}
Y_2(z)=\dfrac{e^{i \delta_3 z}}{ia} \sum_{k=1}^4 C_k \bigg[\alpha_k e^{\xi_1 z} + |b|^2 F(z, \xi_1)\bigg],
\end{multline}
\begin{equation}
Y_3(z)=ib\sum_{k=1}^4 C_k F(z, \xi_1),
\end{equation}
\begin{multline}
Y_4(z)=-\dfrac{c^*}{a}\sum_{k=1}^4 C_k\bigg[\bigg(\alpha_k + \dfrac{|b|^2}{\xi_1}\bigg) F (z,\xi_2) - \\-
\dfrac{|b|^2}{\xi_1}F(z, \delta_4)\bigg],
\end{multline}
where $F(z, \gamma) \equiv (e^{\gamma z} - 1) / \gamma$, 
$\xi_1 = \alpha_k -i\Delta_2$, 
$\xi_2 = \alpha_k - i\Delta_1 +i\Delta_3$, 
$\delta_3 = \Delta_2 - \Delta_1$, 
$\delta_4 = i\Delta_2 + i\Delta_3 -i\Delta_1$.

The Bogoliubov functions are determined in the following way $U_{s}(z)=Y_1(z), \ V_{i}^*(z)=Y_2(z), \ K_{s}(z)=Y_3(z), \ L_{i}^*(z)=Y_4(z)$, with the coefficients $a=\kappa$, $b=\eta_s$, $c=\eta_i$, $\Delta_1=\Delta$, $\Delta_2=\Delta_s$, $\Delta_3=\Delta_i$.

The Bogoliubov functions for the replaced lower indexes are determined in the following way $U_{i}(z)=Y_1(z), \ V_{s}^*(z)=Y_2(z), \ K_{i}(z)=Y_3(z), \ L_{s}^*(z)=Y_4(z)$, with the coefficients $a=\kappa$, $b=\eta_i$, $c=\eta_s$, $\Delta_1=\Delta$, $\Delta_2=\Delta_i$, $\Delta_3=\Delta_s$.

\subsection{Analytical solution for Bogoliubov functions $W(z),\ Q(z),\ M(z),\ N(z)$ }

In the same manner as in previous subsection the functions $W(z),\ Q(z),\ M(z),\ N(z)$ can be found from the system~\eqref{eq_ap_start} with the initial conditions $Y_1(z)=0$, $Y_2(z)=0$, $Y_3(z)=1$, $Y_4(z)=0$.

In this case the~\eqref{eq_ap_general_form} the solution has the form
\begin{equation}
    Y_1(z) = \sum_k D_k e^{ \alpha_k z},
\end{equation}
where $\alpha_k \equiv \gamma_k + i\delta$.
The coefficients $D_k$ are determined by the equation
\begin{equation}
\begin{pmatrix}  1 & 1 & 1& 1 \\ \alpha_1& \alpha_2 & \alpha_3 &\alpha_4 \\ \alpha^2_1& \alpha^2_2 & \alpha^2_3 &\alpha^2_4 \\ \alpha^3_1& \alpha^3_2 & \alpha^3_3 &\alpha^3_4 \end{pmatrix} \begin{pmatrix} D_1 \\ D_2 \\ D_3 \\ D_4 \end{pmatrix}=  \begin{pmatrix} 0 \\ ib^* \\ -b^*\Delta_2 \\  ib^*(|a|^2 - |b|^2 - \Delta^2_2) \end{pmatrix}
\end{equation}
that was obtained in the same way as in previous subsection.

The explicit form of the functions is

\begin{multline}
    Y_2(z)= \dfrac{e^{i \delta_3 z}}{ia}\bigg(-ib^* + \sum_{k=1}^4 D_k\bigg[\alpha_k e^{\xi_1 z} + |b|^2 F(z, \xi_1)\bigg]\bigg)
\end{multline}

\begin{equation}
Y_3(z)=1 + ib\sum_{k=1}^4 D_k F(z,\xi_1)
\end{equation}
\begin{multline}
    Y_4(z)= - \dfrac{c^*}{a}\bigg[\sum_{k=1}^4 D_k \bigg(\alpha_k + \dfrac{|b|^2}{\xi_1}\bigg) F(z, \xi_2) -\\- \bigg(\sum_{k=1}^4 \dfrac{|b|^2 D_k}{\xi_1} + ib^*\bigg) F(z, \delta_4)  \bigg] 
\end{multline}

The Bogoliubov functions are determined in the following way: $W_{s}(z)=Y_1(z), \ Q_{i}^*(z)=Y_2(z), \ M_{s}(z)=Y_3(z), \ N_{i}^*(z)=Y_4(z)$, with the coefficients $a=\kappa$, $b=\eta_s$, $c=\eta_i$, $\Delta_1=\Delta$, $\Delta_2=\Delta_s$, $\Delta_3=\Delta_i$.

The Bogoliubov functions with the replaced lower indexes are: $W_{i}(z)=Y_1(z), \ Q_{s}^*(z)=Y_2(z), \ M_{i}(z)=Y_3(z), \ N_{s}^*(z)=Y_4(z)$, with the coefficients $a=\kappa$, $b=\eta_i$, $c=\eta_s$, $\Delta_1=\Delta$, $\Delta_2=\Delta_i$, $\Delta_3=\Delta_s$.

\subsection{Comments on characteristic equation}
\label{appendix_diff_eq_solution_characteristic}

\paragraph{ Interchange of idler and signal modes}
It could seem that the systems~\eqref{eq_func_system_1} and \eqref{eq_func_system_2} for replaced indexes $s\leftrightarrow i$ are determined by completely different characteristic equations.
However, by applying this replacement for the characteristic equation~\eqref{eq_ap_depressed_form} ($b \leftrightarrow c$ and $\Delta_2 \leftrightarrow \Delta_3$) one can obtain that the second characteristic equation has the form  
\begin{equation}
    \bar{\lambda}^4 + P \bar{\lambda}^2 -i Q \bar{\lambda} + R = 0.
    \label{eq_ap_depressed_form_cc}
  \end{equation}
The roots $\bar{\lambda}_i$ and roots for~\eqref{eq_ap_depressed_form} $\lambda_i$ are related as
\begin{equation}
    \bar{\lambda}_i = \lambda^{\ast}_i.
\end{equation}
Consequently, any of characteristic equations~\eqref{eq_ap_depressed_form} or ~\eqref{eq_ap_depressed_form_cc} can  be used for the parametric amplification analysis.

\paragraph{Characteristic equation analysis with complex coefficients}

The consideration of quartic polynomial is conventionally carried out for real coefficients~\cite{Rees_1922}.
Here, in the characteristic equation~\eqref{eq_ap_depressed_form} the imaginary coefficient by the linear term $\lambda$ is present. 
After the replacement $\mu = i\lambda$ the initial equation~\eqref{eq_ap_depressed_form} is reduced to the form with the real coefficients
\begin{equation}
    \mu^4 - P\mu^2 + Q \mu + R = 0.
\end{equation}
From this point the roots analysis performed in Ref.~\cite{Rees_1922} can be exploited.

\bibliography{cascaded_lit.bib}
 
\end{document}